\pdfoutput=1
\def\year{2020}\relax
\documentclass[letterpaper]{article} 
\usepackage{aaai20}  
\usepackage{times}  
\usepackage{helvet} 
\usepackage{courier}  
\usepackage[hyphens]{url}  
\usepackage{graphicx} 
\usepackage{graphicx}  
\usepackage{amsmath}
\usepackage{amsfonts}
\makeatletter\def\@biblabel#1{}\makeatother

\title{Index Tracking with Cardinality Constraints: \\A Stochastic Neural Networks Approach}
\author{Yu Zheng,\textsuperscript{\rm 1,2} Bowei Chen,\textsuperscript{\rm 3} Timothy M. Hospedales,\textsuperscript{\rm 4} Yongxin Yang\textsuperscript{\rm 2,4}\thanks{Corresponding author: yongxin.yang@ed.ac.uk}\\
\textsuperscript{\rm 1}Southwestern University of Finance and Economics\\
\textsuperscript{\rm 2}ArrayStream Technologies Limited\\
\textsuperscript{\rm 3}University of Glasgow\\
\textsuperscript{\rm 4}University of Edinburgh
}
\begin{document}

\maketitle

\begin{abstract}
Partial (replication) index tracking is a popular passive investment strategy. It aims to replicate the performance of a given index by constructing a tracking portfolio which contains some constituents of the index. The tracking error optimisation is quadratic and NP-hard when taking the $\ell_0$ constraint into account so it is usually solved by heuristic methods such as evolutionary algorithms. This paper introduces a simple, efficient and scalable connectionist model as an alternative. We propose a novel reparametrisation method and then solve the optimisation problem with stochastic neural networks. The proposed approach is examined with S\&P 500 index data for more than 10 years and compared with widely used index tracking approaches such as forward and backward selection and the largest market capitalisation methods. The empirical results show our model achieves excellent performance. Compared with the benchmarked models, our model has the lowest tracking error, across a range of  portfolio sizes. Meanwhile it offers comparable performance to the others on secondary criteria such as volatility, Sharpe ratio and maximum drawdown.
\end{abstract}

\section{Introduction}
\label{sec:intro}

The efficient market hypothesis (EMH) is a financial theory primarily proposed by Samuelson~\cite{Samuelson_1965} and Fama~\cite{Fama_1969}. It states that market prices always contain all of the available information, so they reflect what a company is truly worth. According to the EMH, stocks are traded at their fair values. Therefore, it should be impossible for investors to outperform the overall market because they are not able to buy undervalued or sell overvalued stocks. There has been a long debate about whether the EMH holds true in reality. Investors who believe the EMH mainly adopt passive investment strategies to match, as closely as possible, the performance of a specific index such as S\&P 500. Other investors believe that can select stocks expertly to build portfolios that generate consistent excess returns and beat the market. These approaches are called active investment strategies. Both types of approaches offer value to investors. However, in recent decades, there has been a substantial shift in the asset management industry from active to passive investment strategies as many actively managed funds fail to outperform the market~\cite{Barber_2000,Anadu_2018}.

Index tracking (or replication) is a popular passive investment strategy. It attempts to construct a portfolio of stocks or financial products to replicate the performance of an index. The constructed portfolio is called the tracking portfolio and the chosen index is called the benchmark. How well the constructed portfolio is tracking the index is measured by the tracking error, which is the difference between the index performance and the tracking portfolio performance.  

Methods of constructing a tracking portfolio can be divided into two groups: full replication and partial replication. Full replication involves holding all of the stocks that constitute the chosen index, in their respective weights. It is the most intuitive and transparent index tracking approach. Since the tracking portfolio mirrors the index, its returns should closely track the returns of the index in a frictionless market (i.e., a financial market without transaction costs). However, in practice, full replication may not be the most efficient strategy, due to the higher transaction costs involved in holding a large number of index constituents with frequent rebalancing, churn in index members, and illiquid assets~\cite{Strub2018,BenidisFeng2018}. In contrast, partial replication, as the name implies, only invests in a limited number of index constituents. It usually incurs a larger tracking error but involves smaller transaction costs because the tracking portfolio only contains a subset of stocks from the index. Also, it involves rebalances at lower frequency as full replication usually requires daily rebalance. 

Due its obvious advantages, partial replication approaches have been well studied by academics and financial practitioners. These approaches must address two fundamental issues: asset selection and asset allocation. The former answers the question which stocks should be selected into the tracking portfolio while the latter answers the question how much capital should be allocated to each of the selected stocks. The two issues can be solved either sequentially or jointly through optimisation techniques and we provide a short review of related literature later.

In this paper, we discuss a new partial replication method. Specifically, we formulate the index tracking as a regression problem whose objective is to minimise the tracking error and the weights correspond to the amount to invest in each stock. These weights are therefore subject to constraints including: (i) a long-only strategy where each weight is non-negative; (ii) full capital allocation, so all weights must sum to one; and (iii) the $\ell_0$ norm which determines the tracking portfolio size via enforcing weight sparsity. This full problem is NP-hard, so it is usually solved by heuristic methods. For example, evolutionary algorithms have been widely employed in many previous studies~\cite{BeasleyJohn2003,NiWang2013,LiSun2011}. Here we revisit this optimisation problem from a different angle, proposing a novel reparametrisation that enables it to be solved by simple first-order optimisation methods like gradient descent. In our reparametrised optimisation problem, weights are learnt through stochastic neural networks. We evaluate the proposed approach with S\&P 500 index data and compare it with the forward selection, backward selection and  largest market capitalisation methods. Our approach provides superior tracking accuracy compared to all these widely used financial methods, and across different tracking portfolio sizes. On secondary metrics such as volatility, Sharpe ratio and maximum drawdown, our model offers comparable performance to the alternatives.

There are two distinctive features or contributions of our study in this paper. First, the proposed index tracking approach solves a typical partial replication optimisation problem with stochastic neural networks by using a novel reparametrisation method. To the best of our knowledge, this is one of the few studies that discusses the methodologies for this topic from a connectionist perspective and our approach is simple, efficient and scalable compared to commonly used heuristic alternatives. Second, we perform thorough and large scale experiments to validate the proposed model and offer deep insights. For example, we use over a decade of data from for back-testing, from 31 March 2006 to 31 October 2018. This is a much longer horizon than most existing studies in the literature. We also take the transaction costs into account and examine different tracking portfolio sizes. This thorough evaluation confirms that our approach is robust under various real-world investing scenarios.

\section{Related Work}
\label{sec:related_work}

As mentioned earlier, stock selection and capital allocation are the two fundamental issues for partial (replication) index tracking~\cite{GarciaFernando2017}. The former concerns determining which stocks should be  included in the  portfolio while the latter aims to optimally allocate capital among the chosen stocks to minimise the tracking error. Previous studies can be broadly clustered into two groups according to whether these two issues are solved separately or jointly. 

The first group of methods address these issues in two sequential stages. As the second stage is usually formulated as a regression problem, where the optimal weights of the tracking portfolio are determined by using quadratic programming, the major differences between research methods in this group lie in the techniques used for stock selection in the first stage. For example, hierarchical clustering was employed to select stocks from the index constituents which have similar return performances~\cite{FocardiSergio2004,Dose2005}. Stocks were also selected based on their factor replicating ability~\cite{CorielliFrancesco2006}, and co-integration and correlation~\cite{AlexanderCarol2005}. However, methods in this group are sub-optimal in that in that capital allocation and stock selection are carried out separately rather than jointly optimised.

The second group of methods unifies stock selection and capital allocation by adding a sparsity constraint on the portfolio weights into the tracking error so that the two issues can be optimised simultaneously. The $\ell_0$ norm has been widely used as the sparsity constraint to construct a sparse tracking portfolio~\cite{BenidisFeng2018}. However, imposing the $\ell_0$ constraint makes the regularised regression problem NP-hard and requires search heuristics, such as genetic algorithms~\cite{NiWang2013,LiSun2011,GarciaFernando2017}, Tabu search~\cite{GarciaFernando2017}, simulated annealing~\cite{ChangNigel2000,Woodside2011} and transformation \cite{ColemanThomas2006,WangXu2012}. These algorithms are not guaranteed to find the optimal solution, and in many situations the search space grows super-linearly.

The following sparse index tracking studies are worth mentioning. An evolutionary heuristic for index tracking optimisation with the $\ell_0$ constraint was discussed in~\cite{BeasleyJohn2003}. The optimisation problem was also transformed into mixed-integer linear programming and be solved using a standard integer programming solver~\cite{CanakgozBeasley2009}. A combination of the $\ell_0$ norm and the $\ell_2$ norm was studied in~\cite{TakedaAkiko2013}. However, heuristic methods such as genetic algorithms are very unstable in solving such optimisation problem. The $\ell_1$ norm penalty was added to Markowitz mean-variance framework~\cite{Markowitz1952PORTFOLIO} to derive a sparse and stable portfolio~\cite{BrodieJoshua2009}. The combination of the $\ell_1$ norm and the $\ell_2$ norm was discussed to regularise the regression problem for sparse solution. Although the $\ell_1$ norm is applicable to many portfolio construction problems, it has a fatal conflict with other constraints of index tracking, i.e., the long-only strategy and all the weights sum to one. Removing these constraints, particularly,  the latter can make the $\ell_1$ norm functional~\cite{WU2014116} but this is  a rather non-standard solution to the index tracking problem. Another alternative is to use the fractional norm $\ell_p$, where $0<p<1$, which has no conflict with other constraints. However, it is a non-convex relaxation of the $\ell_0$ norm and increase the difficulty of solving the optimisation problem. Recently, a hybrid heuristic algorithm was proposed to solve the non-convex optimisation problem imposed by $\ell_p$ norm~\cite{Fastrich2014}.

Our method is built on  recent developments in training algorithms for neural networks with stochastic and/or non-smooth neurons. Conventional neural network building blocks are differentiable and deterministic functions, e.g., linear maps and smooth activation functions. However, certain  non-conventional elements can be useful. E.g., a binary neuron whose value is sampled from a Bernoulli distribution parametrised by $p$ can effectively represent a ``hard'' gate. The main challenge for this type of stochastic neural network is estimating the gradient of a loss function w.r.t. the parameters behind the discrete distribution (e.g., $p$ in the above example). Estimating gradients from a stochastic function (e.g., draw a sample from a parametrised distribution) has been studied for years in the operations research community \cite{fu2006gradient}, and in the machine learning community, it was popularised by variational auto-encoders \cite{Kingma2014Auto}, where the core idea is to find a sampling process that separates pure noise and distribution parameters. Early research usually focused on continuous distributions (e.g., Gaussian), but recently some viable solutions for discrete distributions have been proposed \cite{Maddison2017Concrete,Jang2017Categorical}. For discrete distributions, we have to further handle hard non-linearities, and several options have been explored in \cite{Bengio2013Estimating}. In this work, we treat this line of research as a general-purpose optimisation tool and demonstrate its effectiveness and efficiency for index tracking with cardinality constraints.

\section{Methodology}
\label{sec:method}

\subsection{Problem Setting}

Index tracking can be formulated as a regression problem,
\begin{equation}
\label{eq:raw_obj}
\underset{w}{\operatorname{min}}~ \|Xw - y\|_2^2
\end{equation}
where $X\in \mathbb{R}^{D\times N}$ are the log-return of assets and $y\in \mathbb{R}^D$ is the target index (benchmark). $D$ is the number of timesteps (e.g., $D = 750$ trading days in three consecutive years), and $N$ is the number of assets (e.g., $N = 500$ stocks). $w\in \mathbb{R}^N$ is the weight of each asset to hold in order to approximate the index $y$.

In this work, we assume a long-only strategy, which means $w_i \ge 0, ~\forall i$, and the capital is always fully utilised, i.e., $\sum_{i }w_i=1$. With these constraints, the objective function in Eq.~\ref{eq:raw_obj} can be rewritten as,
\begin{equation}
\label{eq:cons_obj}
\underset{w\ge \mathbf{0}, \sum_{i }w_i=1}{\operatorname{min}}~ \|Xw - y\|_2^2
\end{equation}
Eq.~\ref{eq:cons_obj} is a non-negative regression problem with sum-to-one constraint, which can efficiently be solved by quadratic programming (QP).

The quadratic form of Eq.~\ref{eq:cons_obj} is,
\begin{equation}
\label{eq:form}
\begin{split}
\min_w ~~ & \frac{1}{2} w^T P w + q^T w\\
\operatorname{subject~ to:~} & Gw \le h ~\text{and}~ Aw = b
\end{split}
\end{equation}
where $P=2(X^T X)$, $q = - 2X^T Y$, $G=-I$, $h=\mathbf{0}$, $A=\mathbf{1}^T$, and $b=1$. This is a convex optimization problem that can be handled by most of the off-the-shelf solvers.

\subsection{Partial Replication Asset Selection}

The final but most important constraint in partial index replication is that we can only buy \emph{up to} $K$ assets. This reduces transaction costs compared to the full index, and it makes the portfolio more manageable. Thus, the final objective function can be written as,
\begin{equation}
\label{eq:final_obj}
\underset{w\ge \mathbf{0}, \sum_{i }w_i=1, \|w\|_0\le K}{\operatorname{min}}~ \|Xw - y\|_2^2
\end{equation}
$\|w\|_0$ is the $\ell_0$ norm, which is defined as the number of non-zero elements in $w$. Eq.~\ref{eq:final_obj} is much harder to optimise compared to Eq.~\ref{eq:cons_obj} due to the $\ell_0$ norm. In fact, the problem with $\ell_0$ norm has been proven to be NP-hard \cite{Nesterov1994Interior}. This kind of optimisation problem is usually solved by the heuristic methods such as evolutionary algorithms. 

The main focus of this study is to propose a novel reparametrisation for Eq.~\ref{eq:final_obj}, such that it can be solved by plain first-order optimisation methods, e.g., gradient descent. The core of the proposed method is a stochastic asset selection process that models the sparsity in $w$.

\noindent\textbf{Selection Process}\quad
The objective is to select at most $K$ assets to hold from the full index of $N$ assets (e.g., $K=40$ of $N=500$). To better understand the proposed approach, we can imagine that we have $K$ bags, and every bag has $N$ balls corresponding to stocks. We model selection of assets as drawing balls from bags, where the probability of drawing a ball indexed by $j$ from bag $i$ is $\pi_{i,j}$.  These $\pi_{i,j}$'s are auxiliary variables, and we will discuss how learning them can help the optimisation of $w$.

Asset selection can then be described by a stochastic process: we pick exactly one ball from each bag in turn, and take the note of the ball's index every time. This guarantees that we have $K$ or fewer unique indices in the end, and those are the indices of assets that we will buy. 

For the aforementioned stochastic process, we use a one-hot encoding vector to record the outcome of picking a ball from a bag. 
\begin{equation}
\label{eq:sample}
z_i \sim \operatorname{Discrete}([\pi_{i,1},\dots,\pi_{i,N}])
\end{equation}
E.g., for the first bag, if the $5$th ball is picked, we have a vector $z_1=[0,0,0,0,1,0,\dots]$, i.e., in $z_1$, the $5$th element is $1$, and all the remaining are zeros. In the end of the process, we have $K$ index vectors: $\{z_1, z_2, \dots, z_K\}$.

If we take a sum of $z_i$'s, we will have a vector of length $N$: $z=\sum_{i=1}^{K} z_i$, in which at most $K$ elements are non-zeros. $z$ can be seen as a mask vector for assets, only if one element is $1$ or more, the corresponding asset is selected.

\subsection{Reparametrisation}

Optimisation of $\pi$ and $w$ is complicated by the constraints they must meet. Therefore we perform reparametrisation to express them in terms of unconstrained counterparts. 

\noindent\textbf{Auxiliary Asset Selection Probabilities}\quad
The auxiliary variables $\pi_{i,j}$, are all bounded in $[0,1]$. We can generate these in terms of unconstrained parameters $S\in\mathbb{R}^{K\times N}$. For every bag, its probability vector $[\pi_{i,1},\pi_{i,2},\dots,\pi_{i,N}]$ is produced by applying a softmax function on every row of $S$, 
\begin{equation}
\label{eq:pi_wo_annealing}
\pi_{i,j}=\frac{\exp(S_{i,j})}{\sum_j \exp(S_{i,j})}.
\end{equation}

\noindent\textbf{Capital Allocation Weights}\quad
With these preparations, we can now generate capital allocation weights $w$ given a set of unbounded parameters $\tilde{w}\in \mathbb{R}^N$ in three steps:

\begin{enumerate}
\item Element-wise exponentiation ensures positivity: $\hat{w}=[\exp{(\tilde{w}_1)}, \exp{(\tilde{w}_2)},\dots,\exp{(\tilde{w}_N)}]$
\item Element-wise product with $z$ allocates capital only to picked stocks: $\bar{w}=\hat{w}\odot z$.
\item Normalisation $w=\frac{\bar{w}}{\sum_i \bar{w}_i}$ ensures all capital is allocated.
\end{enumerate}
After those steps, we can easily verify the capital allocation regression weights $w$ meet all three conditions for partial replication tracking: $w\ge \mathbf{0}, \sum_{i }w_i=1, \|w\|_0\le K$.

\subsection{Objective Function}
With the reparametrisation derived above, the objective function in Eq.~\ref{eq:final_obj} can be written as,
\begin{equation}
\label{eq:alter_obj}
\underset{\tilde{w}, S}{\operatorname{min}}~ \|Xw - y\|_2^2
\end{equation}
where the parameters in the optimisation problem are all now unbound, i.e., $\tilde{w}\in \mathbb{R}^N$ and $S\in\mathbb{R}^{K\times N}$. Here $w$ is a stochastic function $w=f(\tilde{w}, S, \epsilon)$ and $\epsilon$ stands for the randomness introduced by the sampling process (picking balls from bags). 

Even though $\tilde{w}$ and $S$ are unbounded now, there is still one issue that stops us from using gradient-based optimisation: the sampling process used for portfolio selection (Eq.~\ref{eq:sample}). The partial replication generator $w=f(\tilde{w}, S, \epsilon)$ can be understood as a \emph{stochastic} neural network with \emph{discrete} hidden neurons $z$, which is hard to train because the backpropagation algorithm can not be applied to the non-differentiable operations including the sampling function used here. Thanks to the recent development of stochastic gradient estimators \cite{Kingma2014Auto}, we can get a low-variance estimator of the gradients as discussed next.

\subsection{Gradient Estimation}

The key ingredient to solve the gradient estimation issues is the so-called Gumbel-Softmax trick \cite{Maddison2017Concrete,Jang2017Categorical}, along with Straight Through (ST) gradient estimator \cite{Bengio2013Estimating}.

Instead of drawing a sample $z$ from the categorical distribution parametrised by $[\pi_1,\pi_2,\dots,\pi_N]$ directly, we use the Gumbel-Max trick \cite{gumbel1954statistical},
\begin{equation}
\label{eq:gumbelmax}
z=\operatorname{one\_hot}(\underset{i}{\operatorname{argmax}}(g_i+\log(\pi_i)))
\end{equation}
where $[g_1,g_2,\dots,g_N]$ are i.i.d samples drawn from Gumbel distribution with location $0$ and scale $1$. The sampling process can be realised by first generating a Uniform(0,1) sample $u$ and then calculating $g=-\log(-\log(u))$.

The motivation behind Eq.~\ref{eq:gumbelmax} is to `separate' the trainable variables (i.e., $\pi_i$s) from pure noise (i.e., $g_i$s), so that taking the gradient of random sample with respect to the distribution parameters becomes possible.

Eq.~\ref{eq:gumbelmax} is not sufficient, as $\operatorname{argmax}$ is non-differentiable too. Thus, we use softmax function as a continuous relaxation,
\begin{equation}
\label{eq:gumbelsoftmax}
z=\operatorname{softmax}(g+\log(\pi))
\end{equation}
where $g=[g_1,g_2,\dots,g_N]$ and $\pi=[\pi_1,\pi_2,\dots,\pi_N]$. Eq.~\ref{eq:gumbelsoftmax} is called Gumbel-Softmax estimator in \cite{Maddison2017Concrete,Jang2017Categorical}.

However, it is crucial to make sure $z$ is a \emph{true} one-hot encoding vector, as it will be used for asset selection eventually. As a middle ground, we use Eq.~\ref{eq:gumbelmax} in the forward pass and Eq.~\ref{eq:gumbelsoftmax} in the backward pass. This is known as Straight Through Gumbel-Softmax estimator \cite{Bengio2013Estimating,Jang2017Categorical}.

\subsubsection{Optimisation with Annealing}

The optimisation of Eq.~\ref{eq:alter_obj} is now straightforward, as we can use gradient descent or its variants. Note that, in each forward-backward round, a set of new $z_i$'s will be sampled, so that the algorithm has the opportunity to explore many combinations of assets.  

To guarantee the converge, we introduce an annealing process that progressively reduces the randomness of sampling, such that, for each of the $K$ selections, only the one with the highest probability will be chosen in the end. This can be realised by adding a temperature term into the reparametrisation of $\pi_{i,j}$ in Eq.~\ref{eq:pi_wo_annealing}
\begin{equation}
\label{eq:pi_w_annealing}
\pi_{i,j}(\tau)=\frac{\exp(S_{i,j}/\tau)}{\sum_j \exp(S_{i,j}/\tau)}
\end{equation}
\noindent where $\tau=0.1/\log(\mathrm{e}+t)$ and $t$ is the iteration index.

We should emphasise that the designed annealing process is not the same as described in Gumbel-softmax estimator \cite{Jang2017Categorical}, where the authors added a temperature term into Eq.~\ref{eq:gumbelsoftmax}:
\begin{equation}
\label{eq:gumbelsoftmax_w_tau}
z=\operatorname{softmax}((g+\log(\pi))/\tau)
\end{equation}
\noindent but we found it unnecessary in our case. The reason is that we use the straight-through estimator, which delivers a \emph{true} one-hot vector even with high temperature.

\subsubsection{Post-processing}

The most valuable part of this method is the asset selection, and it is fairly easy to solve the constrained regression problem in Eq.~\ref{eq:cons_obj} if the selection of assets is given. Therefore we can optionally post process the results of our proposed method using conventional constrained regression. The overall procedure is:  (i) Train our model by optimising Eq~\ref{eq:alter_obj}. (ii) Retrieve the selected assets' IDs by taking the index of the maximum of each row in matrix $S$; (iii) Get the unique IDs in (ii); (iv) Use those IDs to select a subset of $X$ such that $X'\in \mathbb{R}^{D\times K'}$ (here $K'$ is the number of selected assets and it should be smaller than or equal to $K$); (v) Solve the constrained regression problem,
\begin{equation}
\underset{w'\ge \mathbf{0}, \sum_{i }w'_i=1}{\operatorname{min}}~ \|X'w' - y\|_2^2
\end{equation}
with Quadratic Programming (QP) using Eq.~\ref{eq:form}.

\begin{figure*}[h!]
\centering
\includegraphics[width=0.32\linewidth]{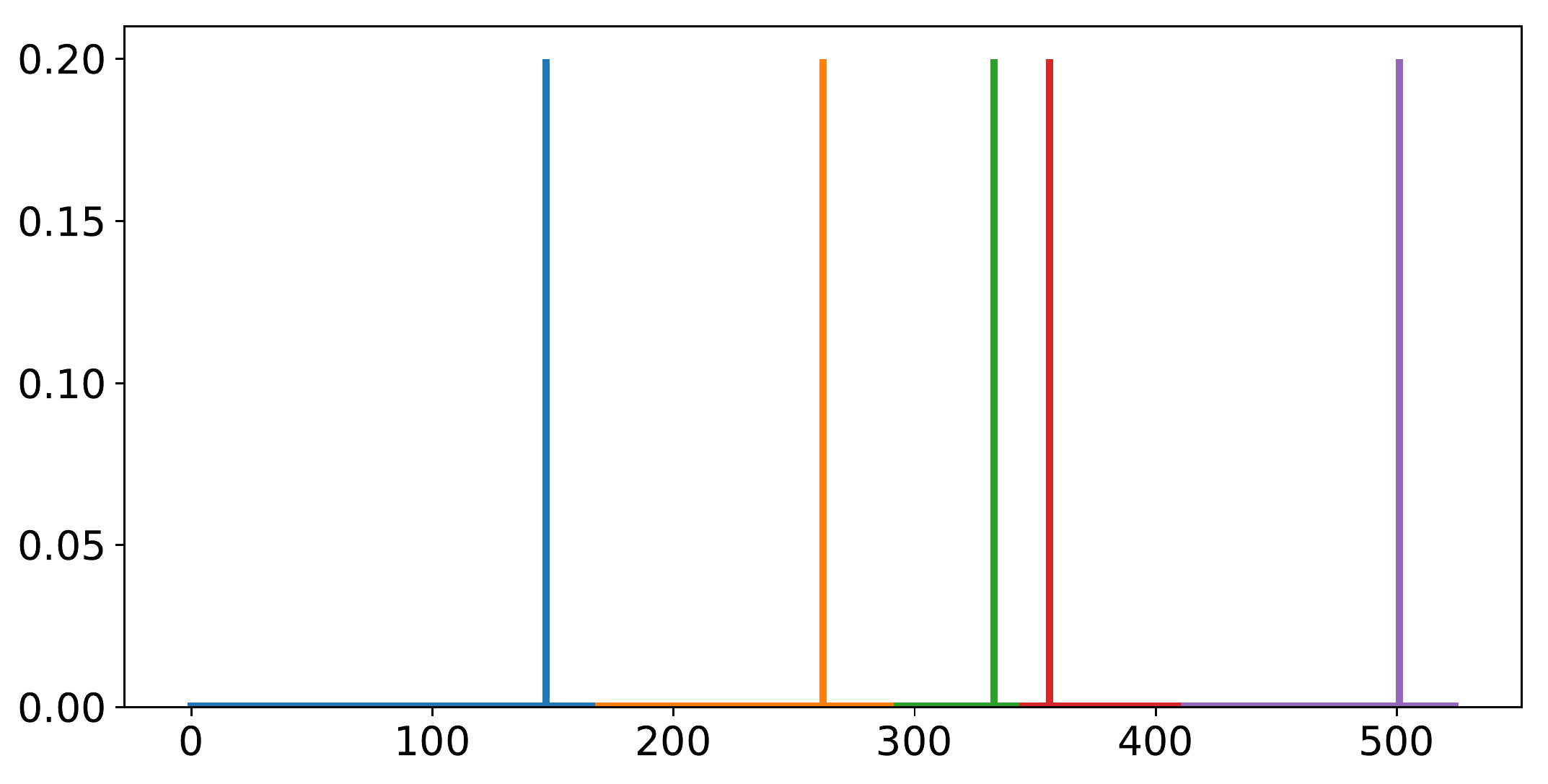}~
\includegraphics[width=0.32\linewidth]{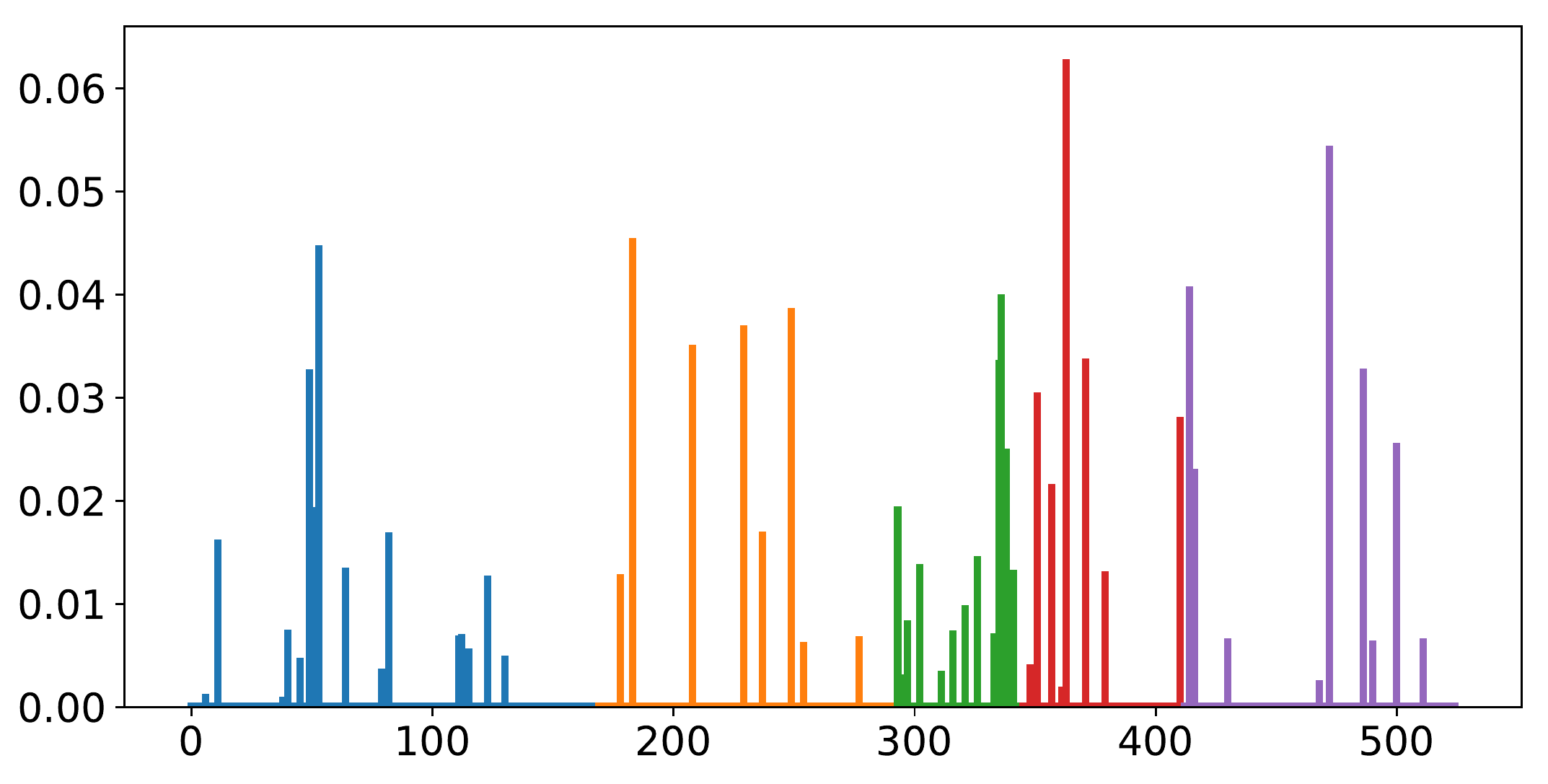}~
\includegraphics[width=0.32\linewidth]{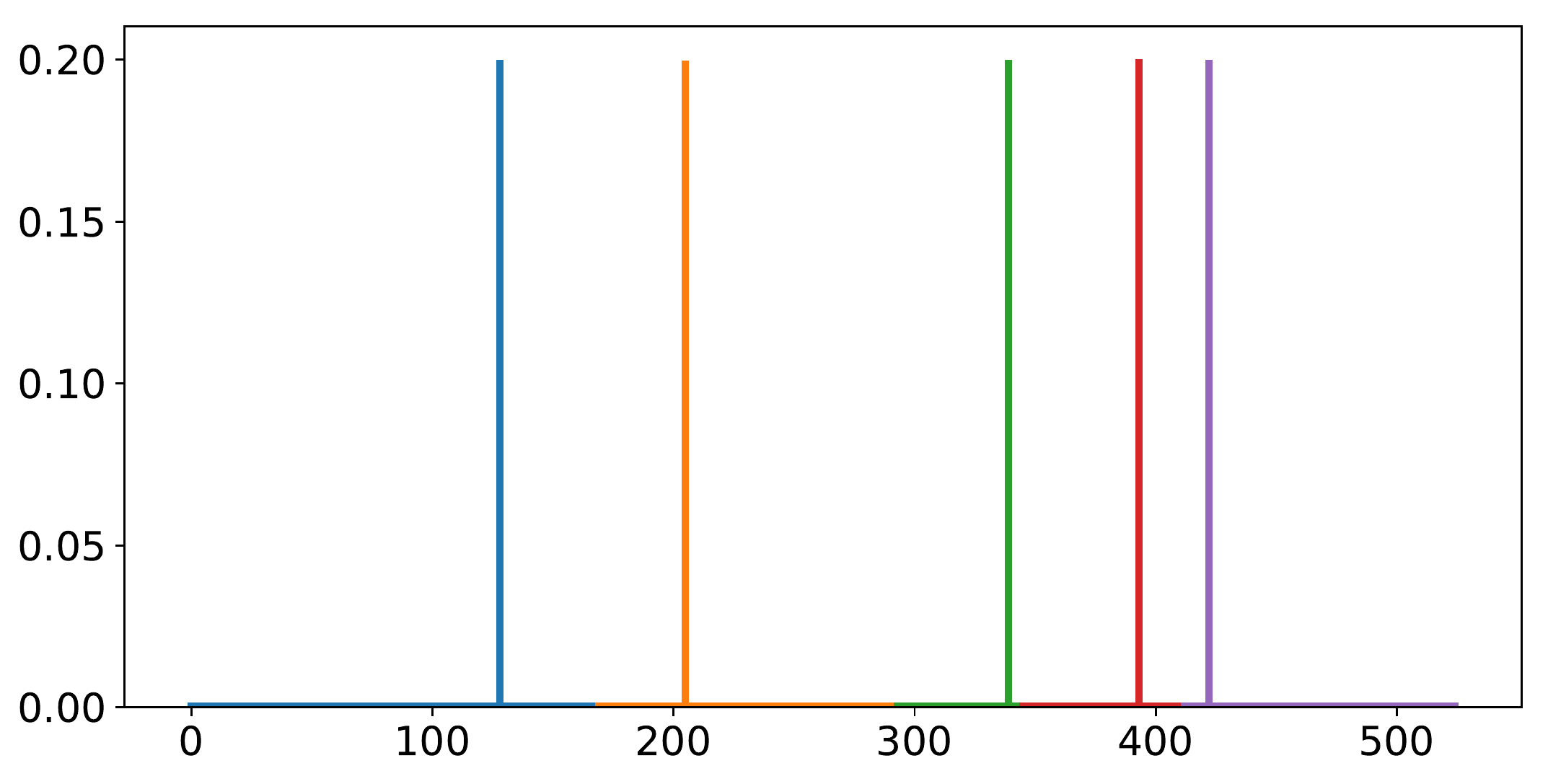}
\caption{Toy asset selection problem, where each colour groups one `sector' of `stocks'. X-axis: 500 possible stocks to hold. Y-axis: Capital allocation to each stock. The ideal solution holds exactly one stock in each sector with equal allocation across sectors. Left: One of many possible ideal solutions. Middle: Solution given by QP with all constraints except for capped $K$. Right: Solution given by our method with $K=5$.}
\label{fig:toy}
\end{figure*}

\section{Experiments}
\label{sec:experiments}

\subsection{Synthetic Example}

We first build a toy problem to illustrate our algorithm. First, we generate $5$ random samples, $\{x_1, x_2, x_3, x_4, x_5\}$, from a multivariate Gaussian distribution, where each $x_i$ is a $750$-dimensional vector. Then we generate the ground truth $Y$ as $Y=0.2x_1 + 0.2x_2 + 0.2x_3 + 0.2x_4 + 0.2x_5$, which can be treated as true (benchmark) index. To generate $X$, each $x_i$ is duplicated by $N_i$ times, where $N_i$ is a random integer in $[50,200]$, then we concatenate those repeated $x_i$'s, i.e., $X=[x_1,x_1,\dots,x_1,x_2,x_2,\dots,x_5,x_5]$. Finally, we add some small Gaussian noise to all entries in $X$ and $Y$. This synthetic problem corresponds to a situation where there are many assets grouped into five underlying classes which determine the overall benchmark.

We can tell that the good solution $w$ to the regression problem $\|Xw-Y\|_2^2$ with non-negativity and sum-to-one constraints should be very sparse. The perfect solution to $w$ should have exactly $5$ non-zero values, each corresponding to one of those $x_i$'s, as illustrated in Fig.~\ref{fig:toy}. Note that Fig.~\ref{fig:toy} (Left) is just one of many possible optimal solutions since each $x_i$ is duplicated, meaning that choosing any of its copies is equally good. 

If we try to solve this constrained regression problem using quadratic programming without cardinality constraints (i.e., Eq.~\ref{eq:form}), we may have a solution with good fitting performance, however, the number of assets used is very large (Fig.~\ref{fig:toy} (Middle)), which results in a high transaction fee. 

In contrast, if $K=5$ is given, our method perfectly selects an equally weighted portfolio of one asset from each of the asset groups (Fig.~\ref{fig:toy} (Right)).

\subsection{S\&P500 Index Tracking}

To rigorously evaluate our method's real-world performance, we track the S\&P500 index using the proposed method and compare it with several baselines.

\subsubsection{Data Preparation}

The pricing data is obtained from the Center for Research in Security Prices (CRSP), which is known to be the most accurate data for study. We use the daily closing prices adjusted for dividends. To simulate real trading, we also take into account the transaction fee, for which we choose the flat-fee pricing model, \$5.00 per trade, quoted by TradeStation (a popular US online stock brokerage firm).

\subsubsection{Backtesting}

Frequent rebalancing can reduce tracking error, however this also leads to high transaction costs. As we target long-term investment, we adopt quarterly portfolio rebalancing for all compared methods. 

The backtesting period is the last decade, starts from 2009-01-02 and ends by 2018-10-31. The backtesting is by sliding window model fitting and evaluation as follows: at the end of each quarter, e.g., on the 31st March 2009, we rebalance the portfolio according to the weights calculated by our method. To be more specific, on the 31st March 2009, we train our model using the data in the last three years (i.e., 2006-03-31 to 2009-03-31), then we buy the stocks suggested by the model ($w$) and hold until the next rebalance day, i.e., 2009-05-31. This procedure is repeated quarterly until the last trading day 2018-10-31. The backtesting simulator accounts for all real-world details including trading costs, stock splits, dividends, index churn, etc.

\begin{figure*}[h!]
\centering
\includegraphics[width=0.35\linewidth]{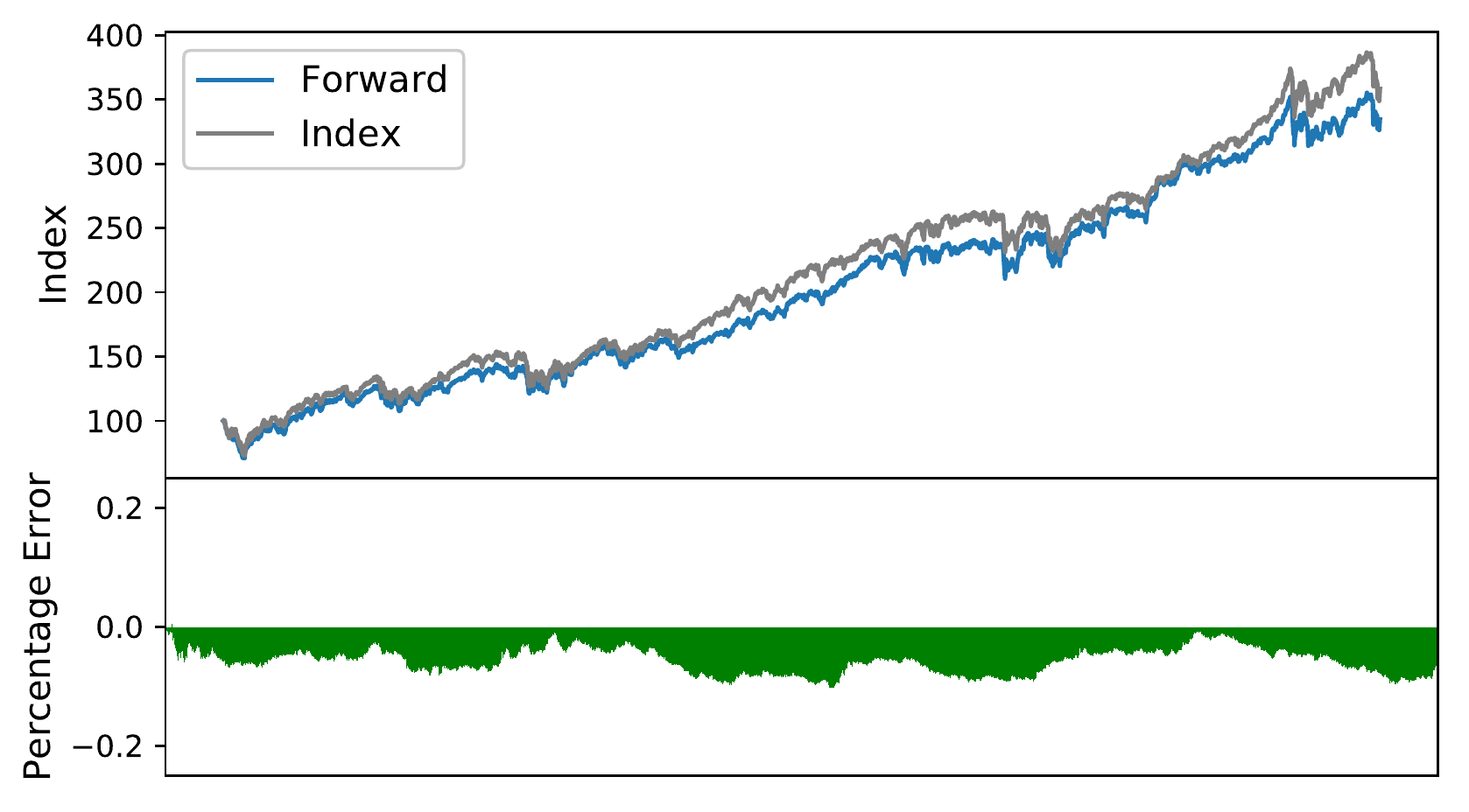}~
\includegraphics[width=0.32\linewidth]{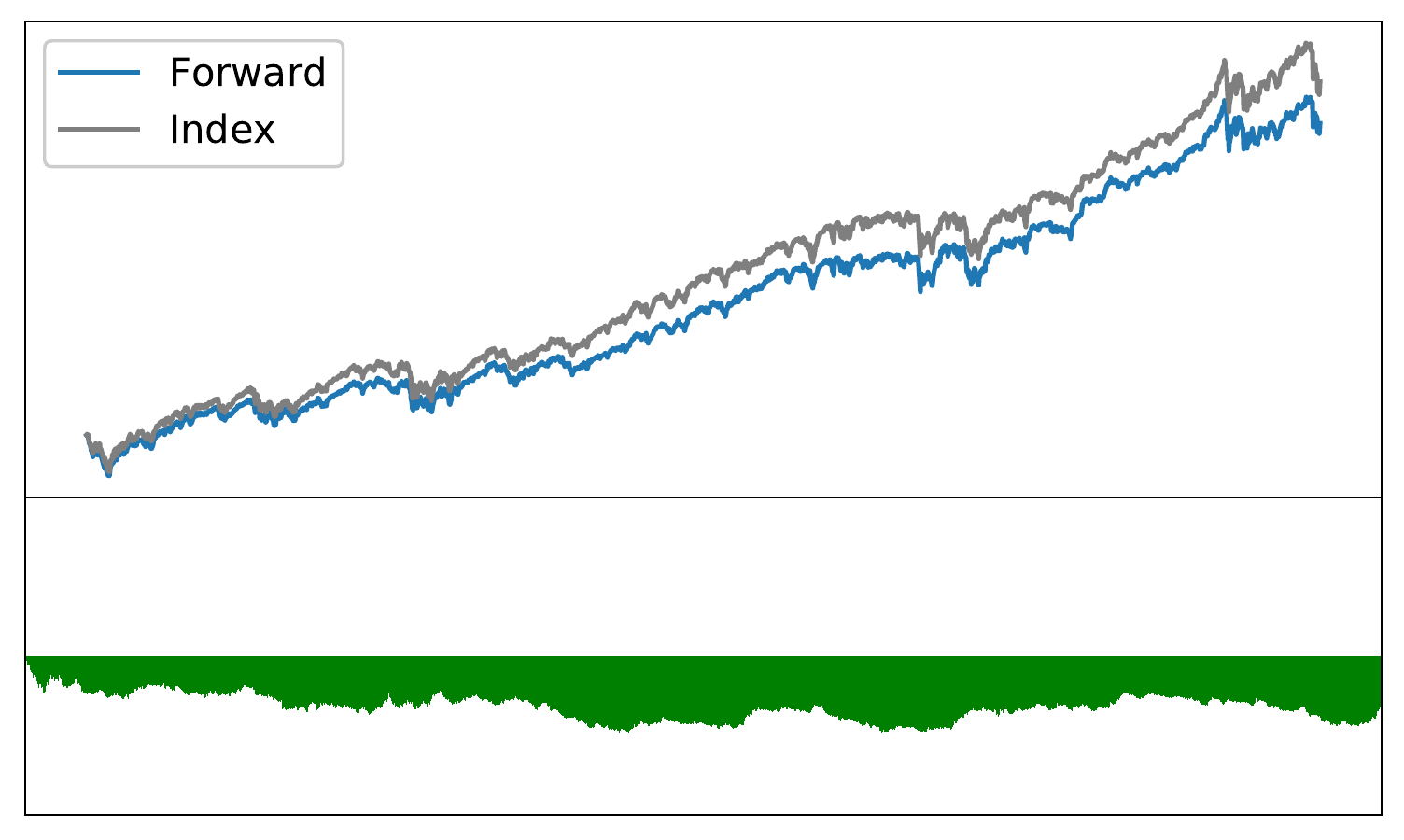}~
\includegraphics[width=0.32\linewidth]{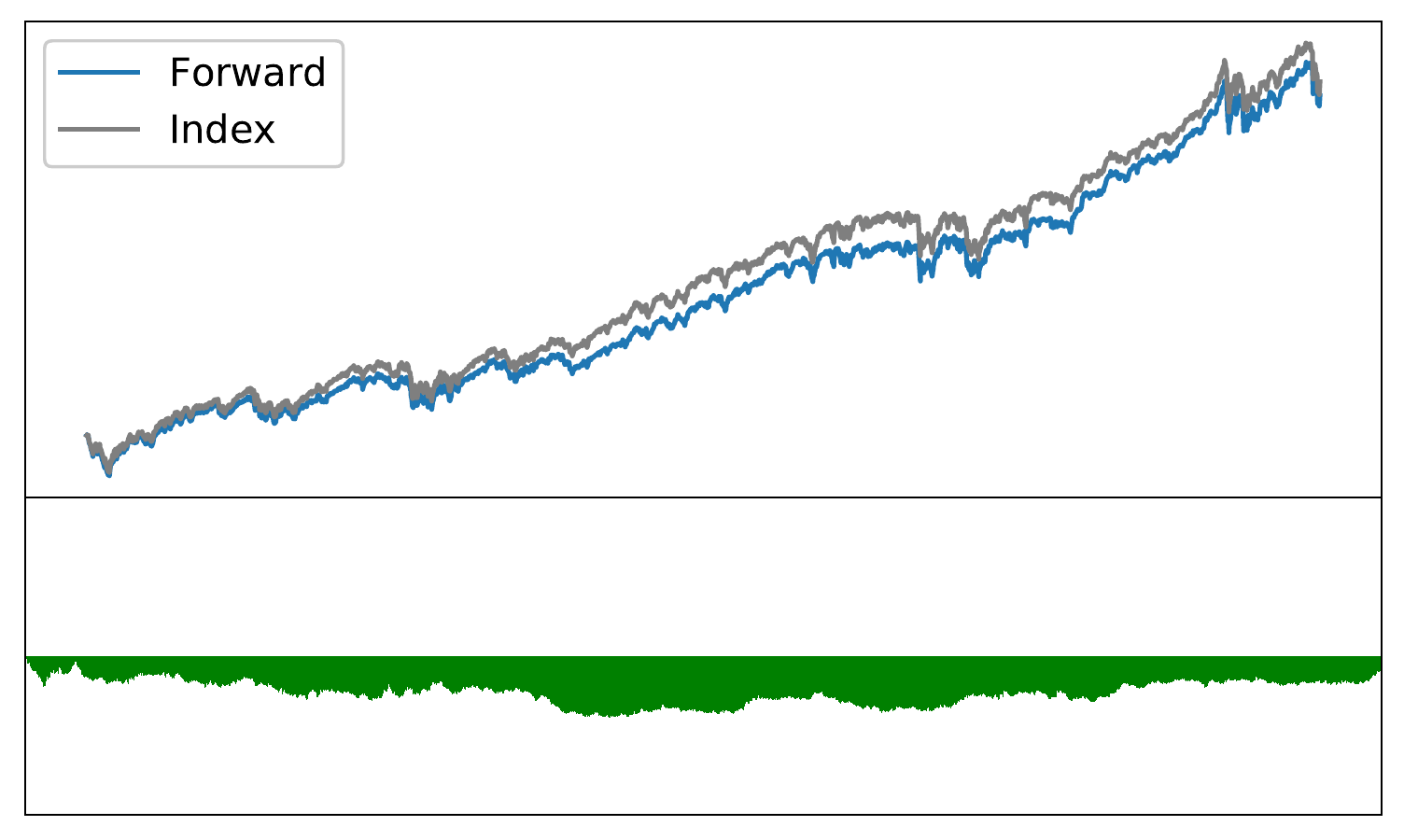}\\
\includegraphics[width=0.35\linewidth]{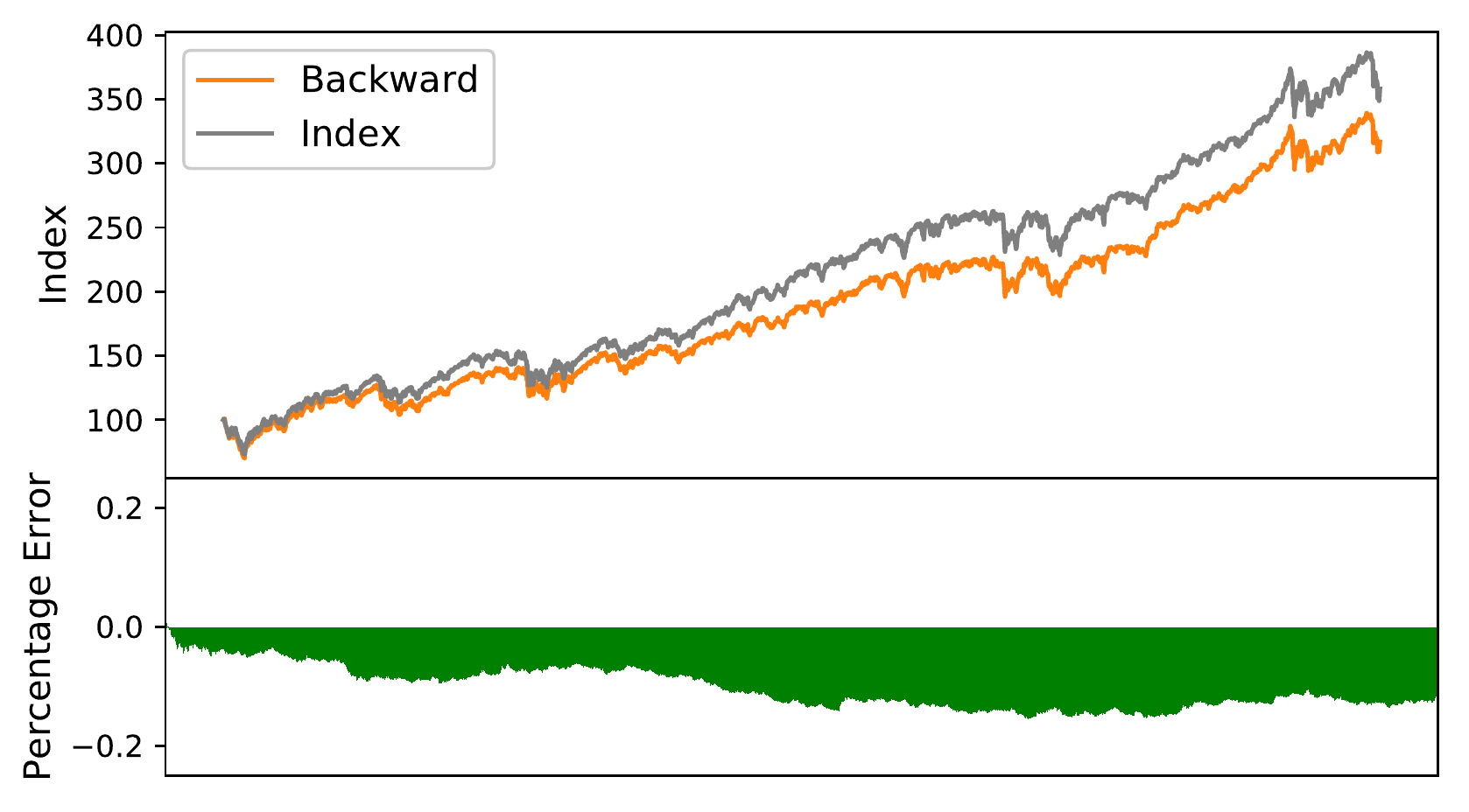}~
\includegraphics[width=0.32\linewidth]{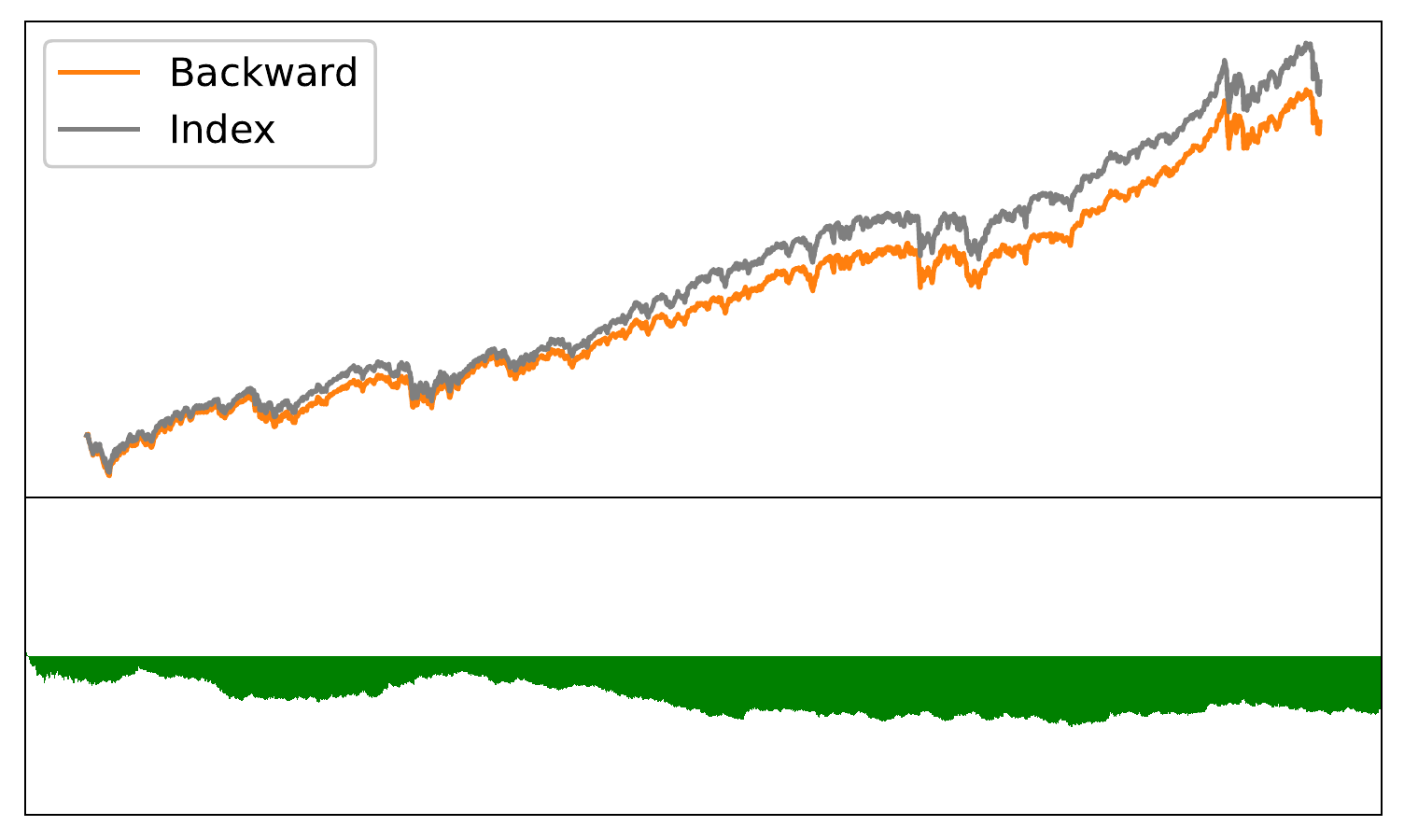}~
\includegraphics[width=0.32\linewidth]{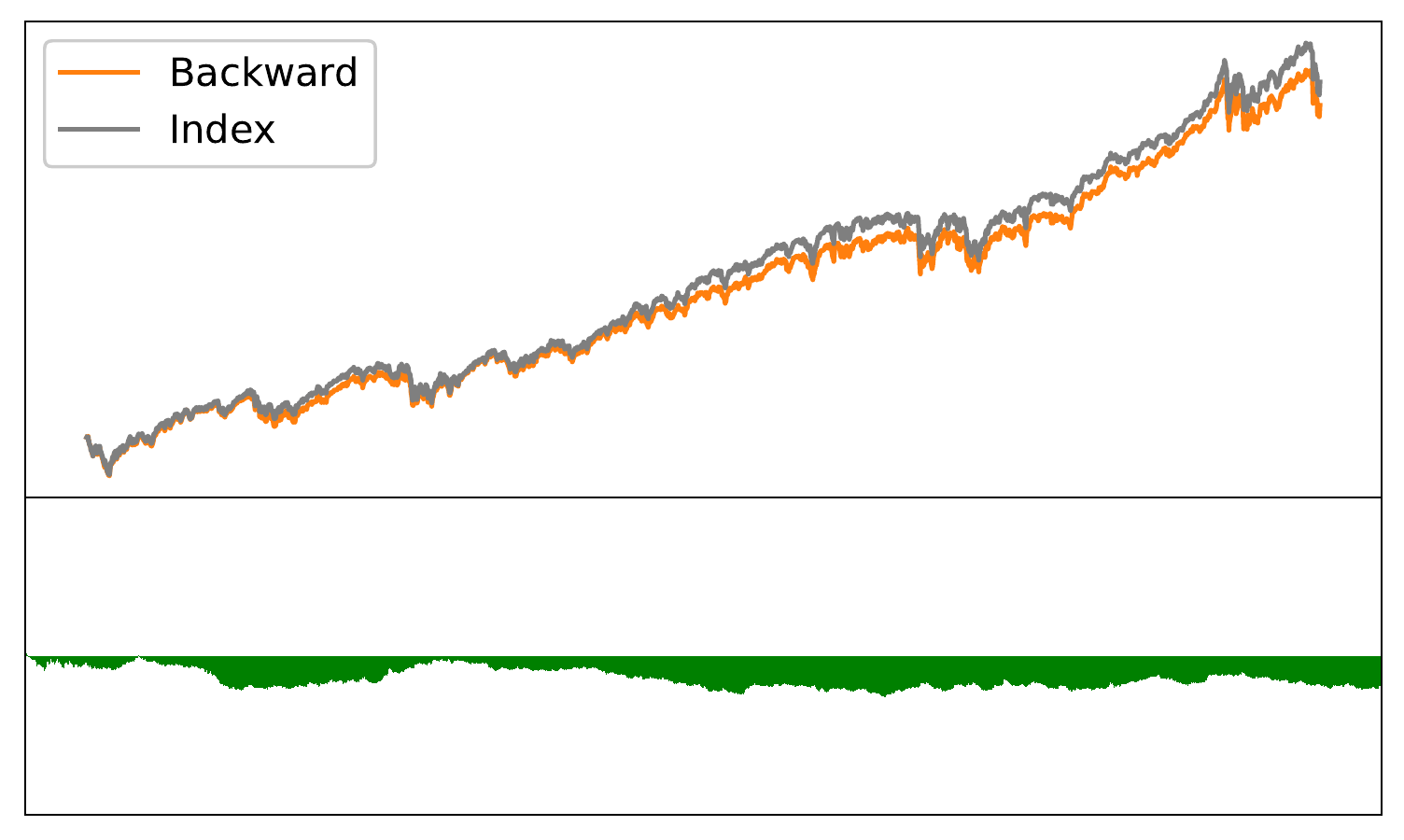}\\
\includegraphics[width=0.35\linewidth]{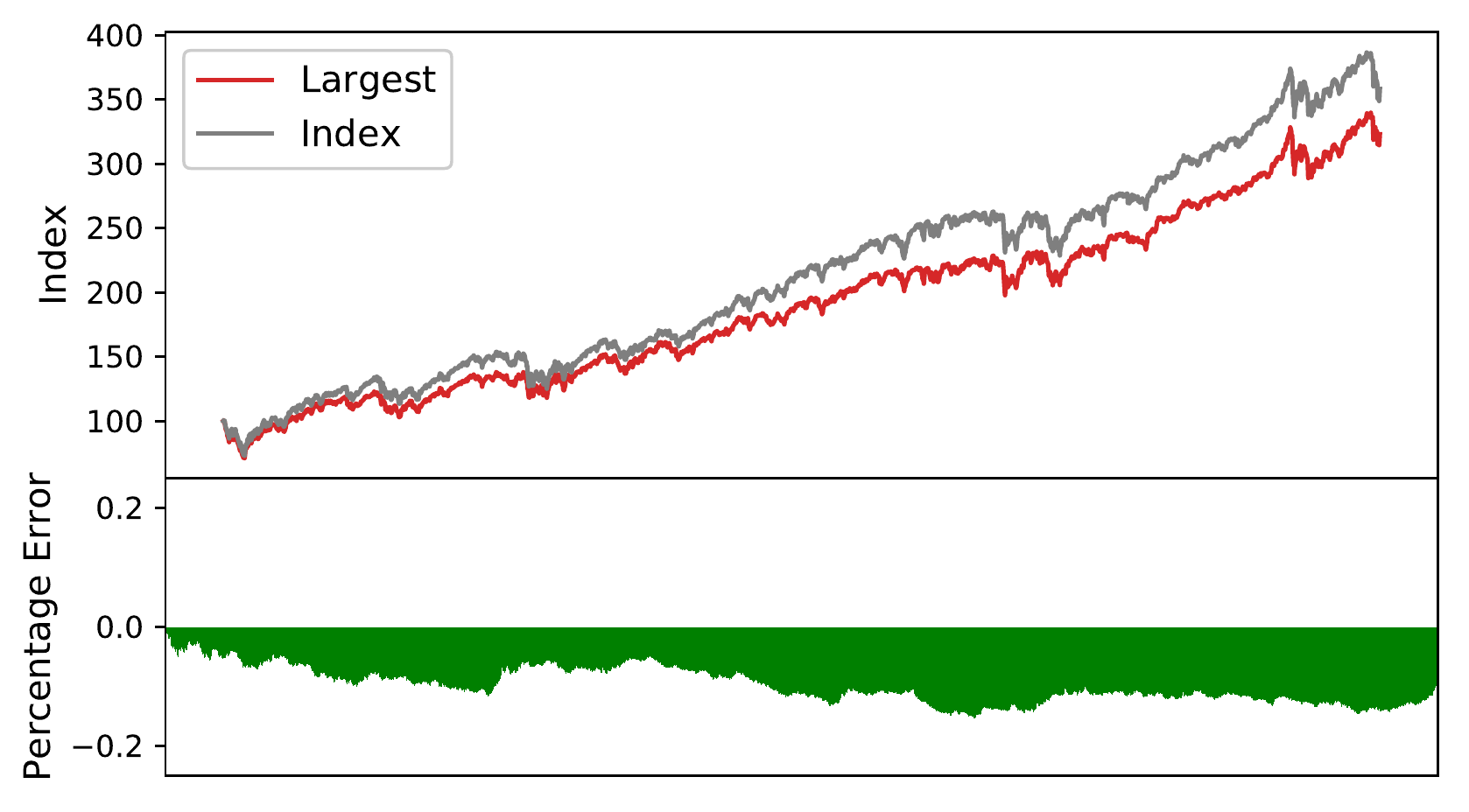}~
\includegraphics[width=0.32\linewidth]{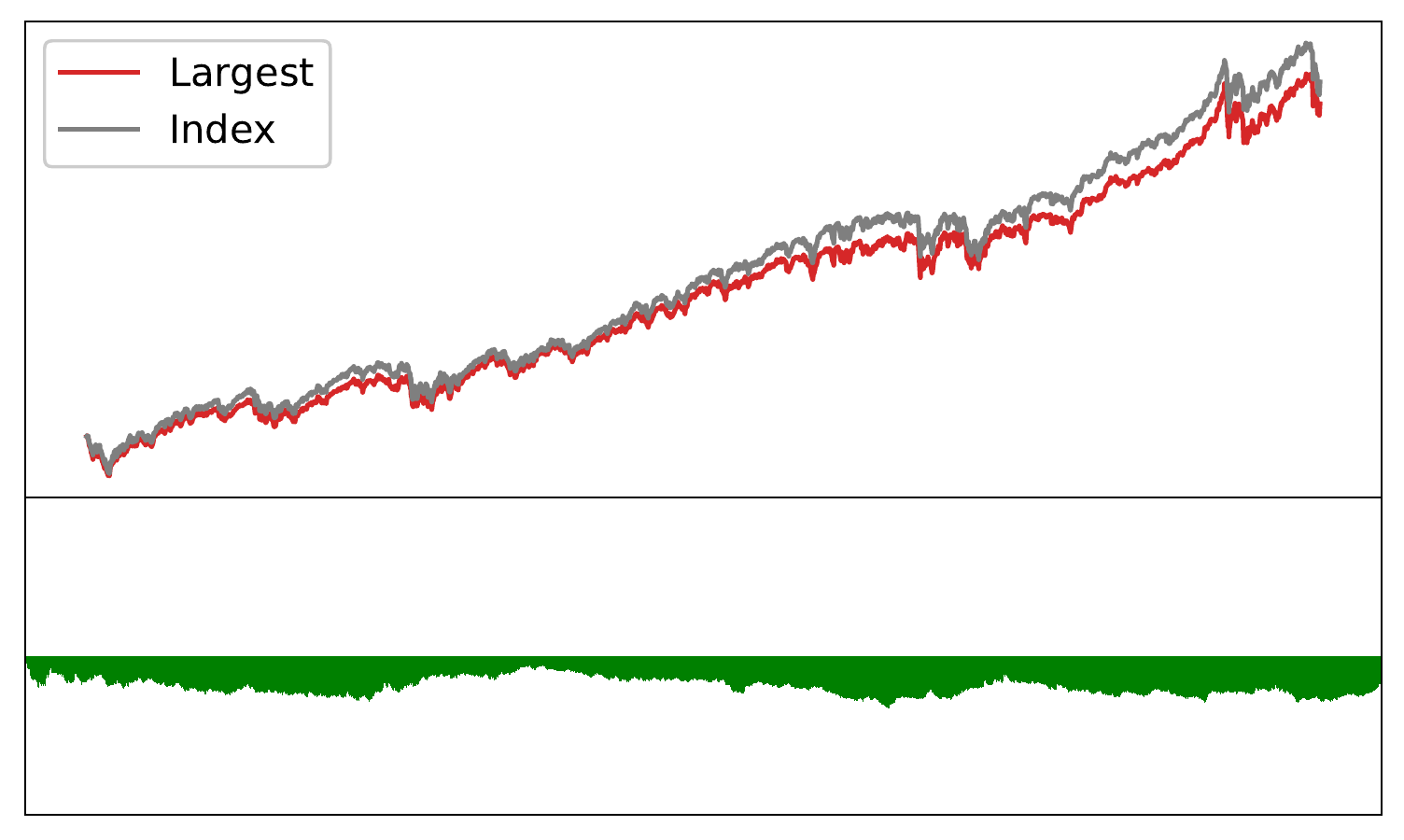}~
\includegraphics[width=0.32\linewidth]{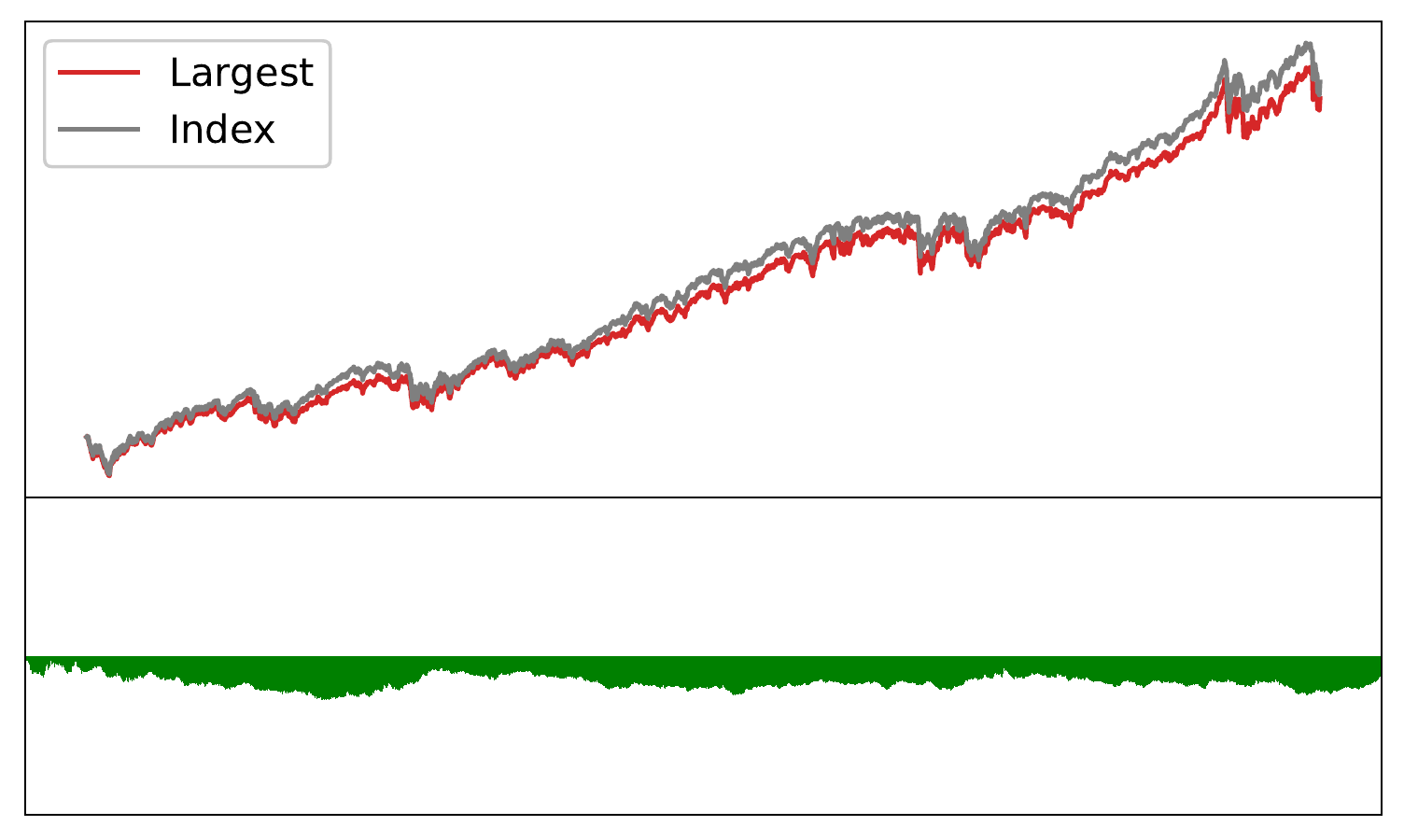}\\
\includegraphics[width=0.35\linewidth]{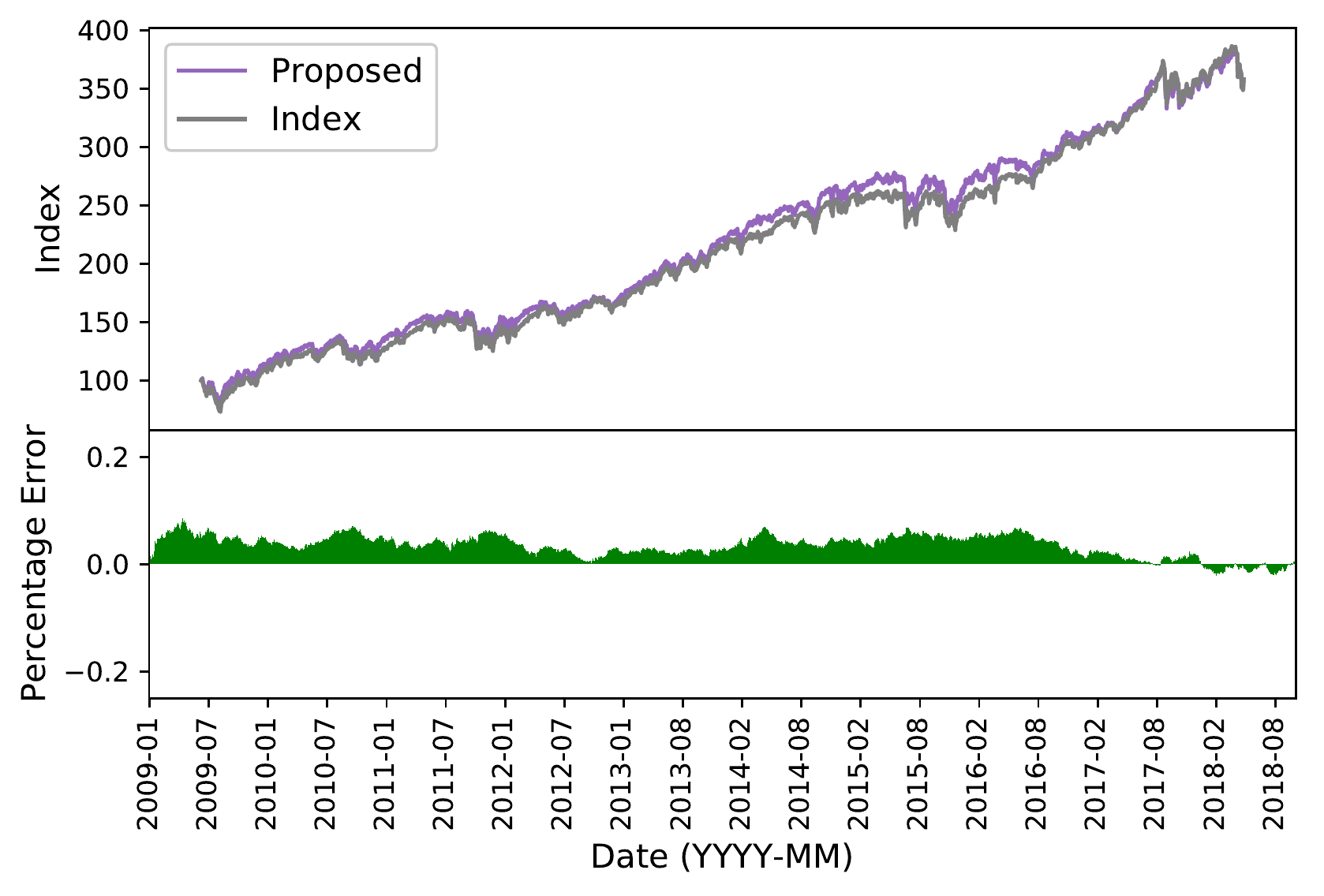}~
\includegraphics[width=0.32\linewidth]{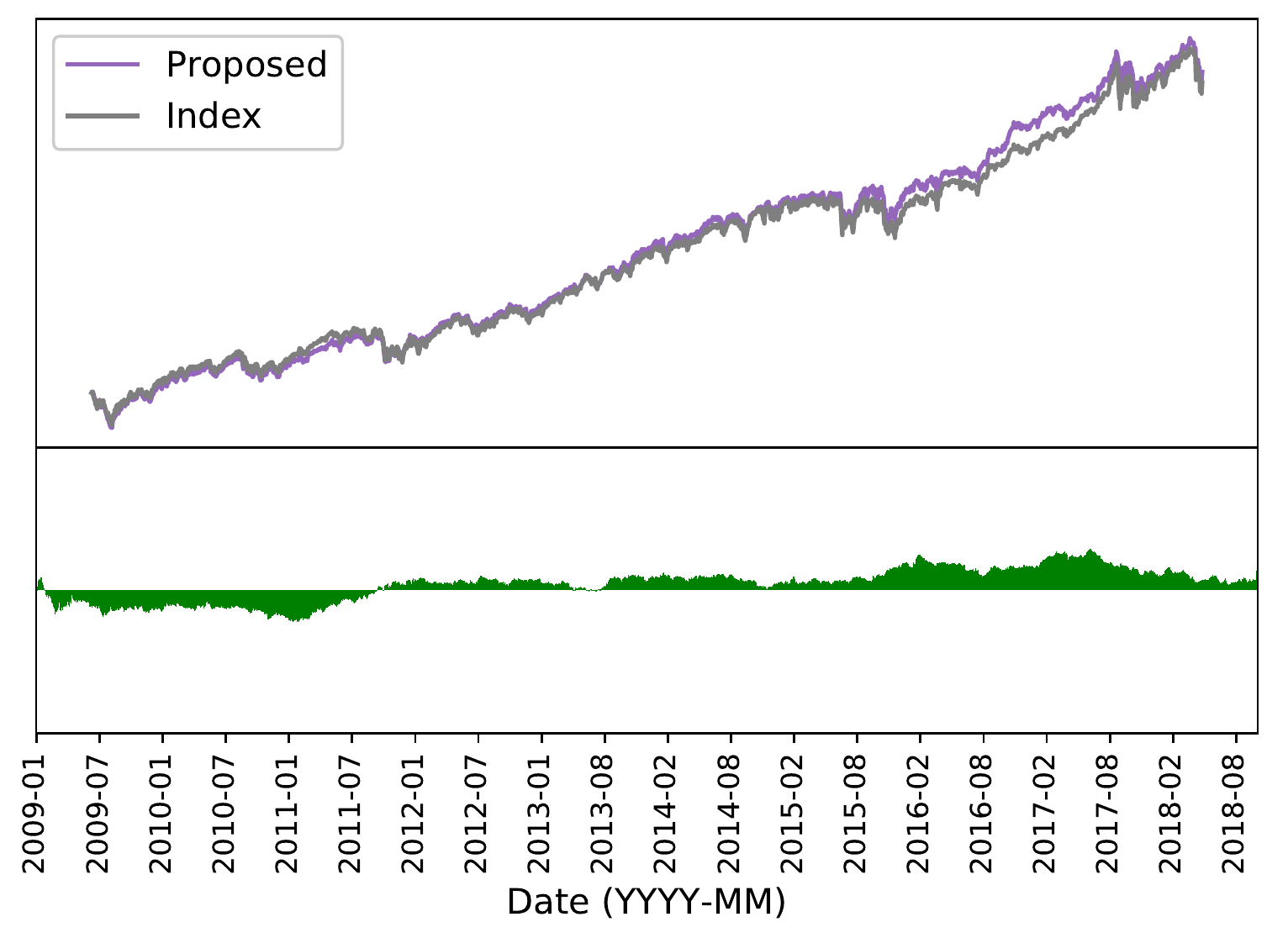}~
\includegraphics[width=0.32\linewidth]{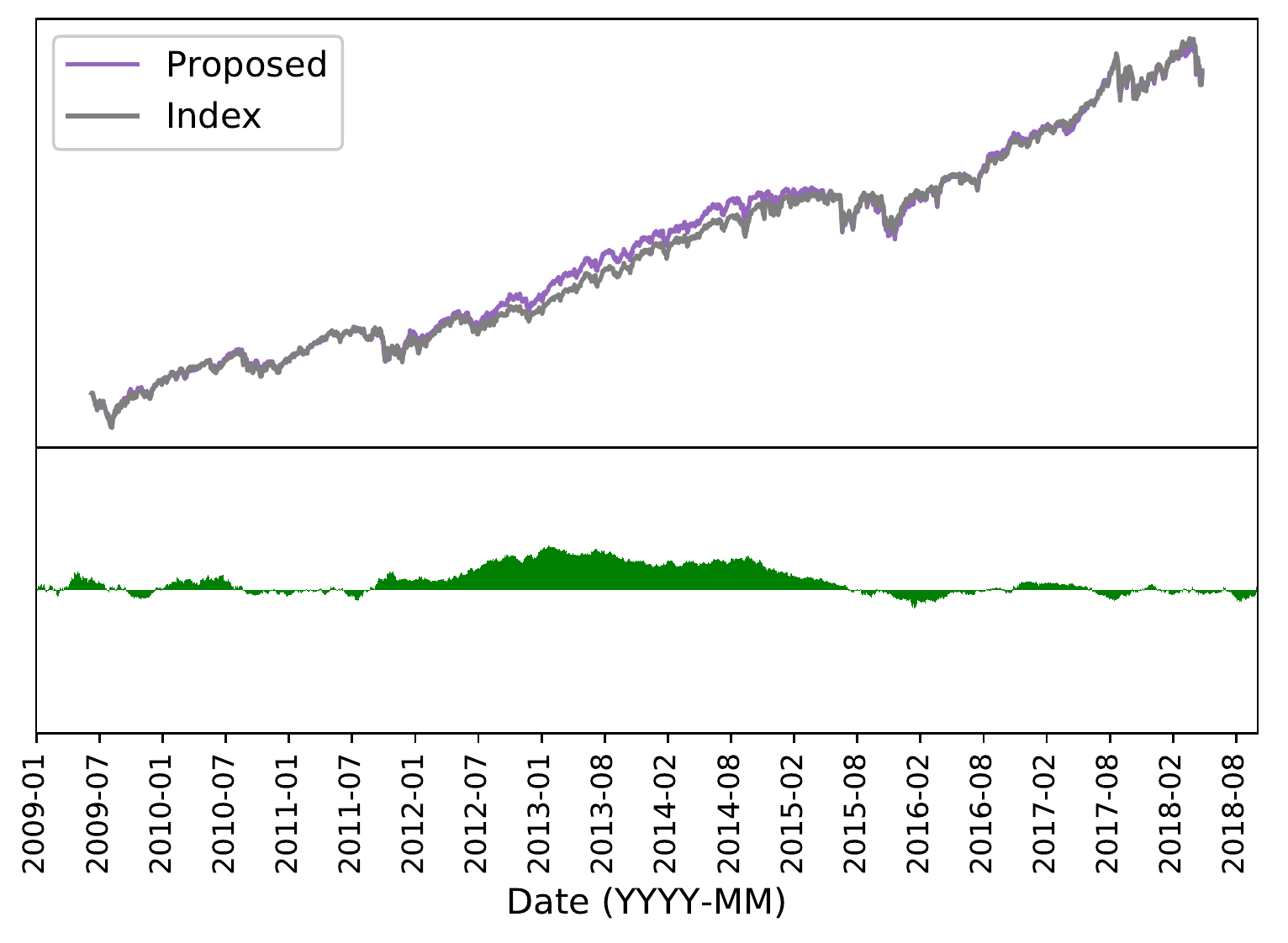}
\caption{Equity curve for the index and trackers. Rows from top to bottom compare different partial replication trackers: Forward-selection, Backward-selection, Largest market cap, Proposed method. Columns evaluate different portfolio sizes $K=[30,40,50]$ from left to right.
In every subplot, the top section (two curves) shows the index (grey) and the tracker, and bottom section (green bars) shows the percentage tracking error $\frac{\hat{y}-y}{y}$.}
\label{fig:fitting}
\end{figure*}

\subsubsection{Performance measure}

\begin{figure*}[h!]
\centering
\includegraphics[width=0.32\textwidth]{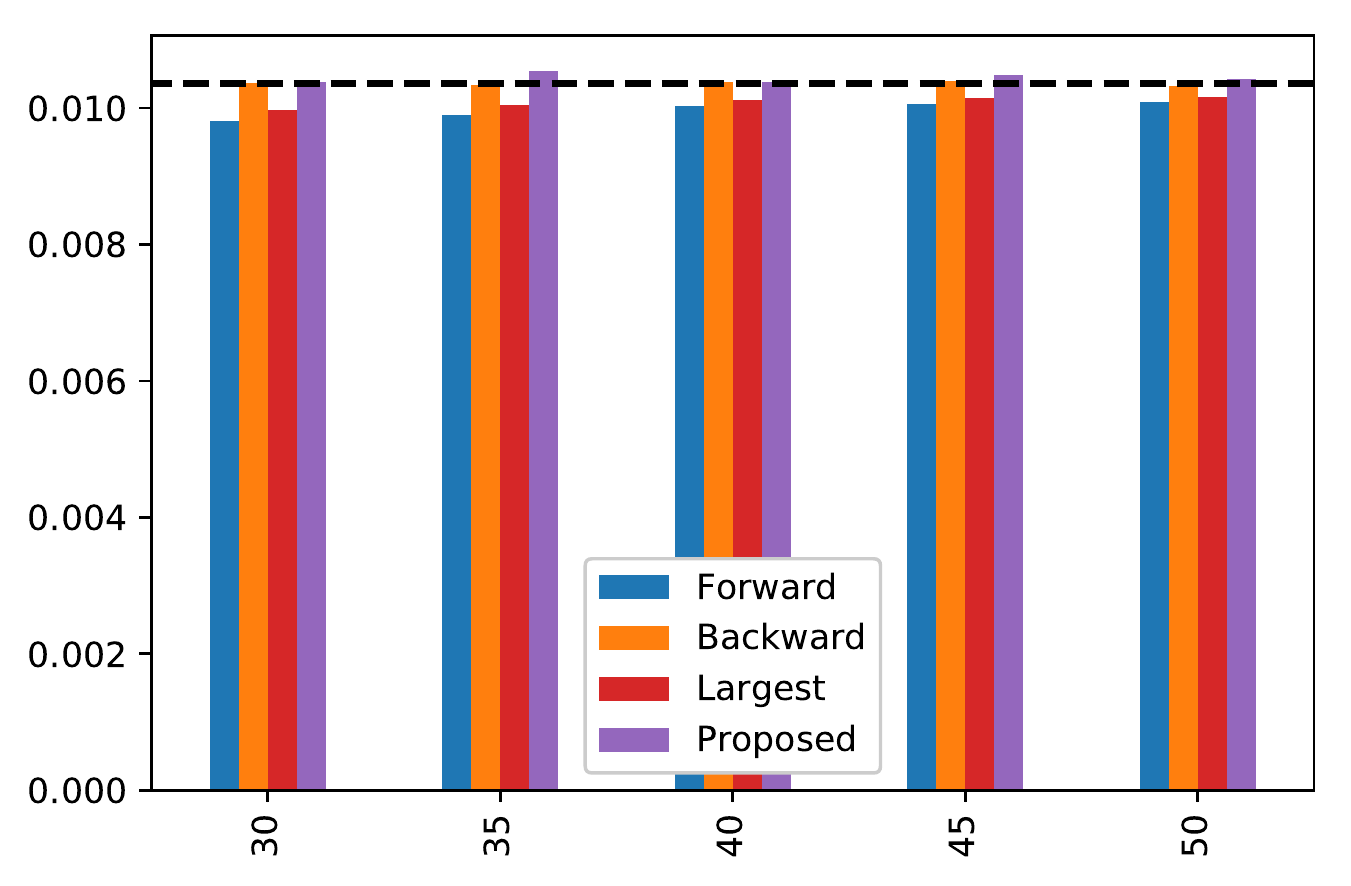}~
\includegraphics[width=0.32\textwidth]{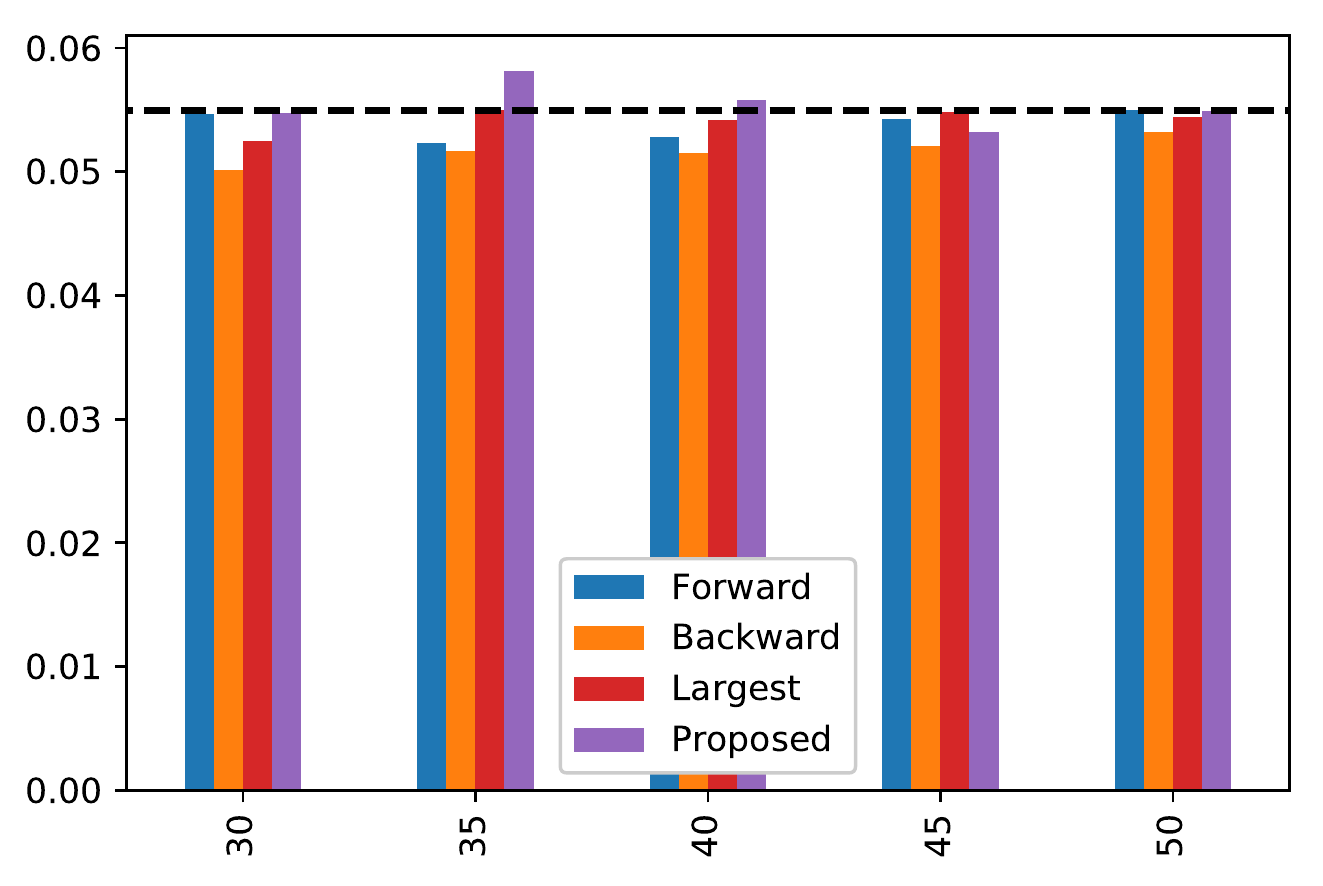}~
\includegraphics[width=0.32\textwidth]{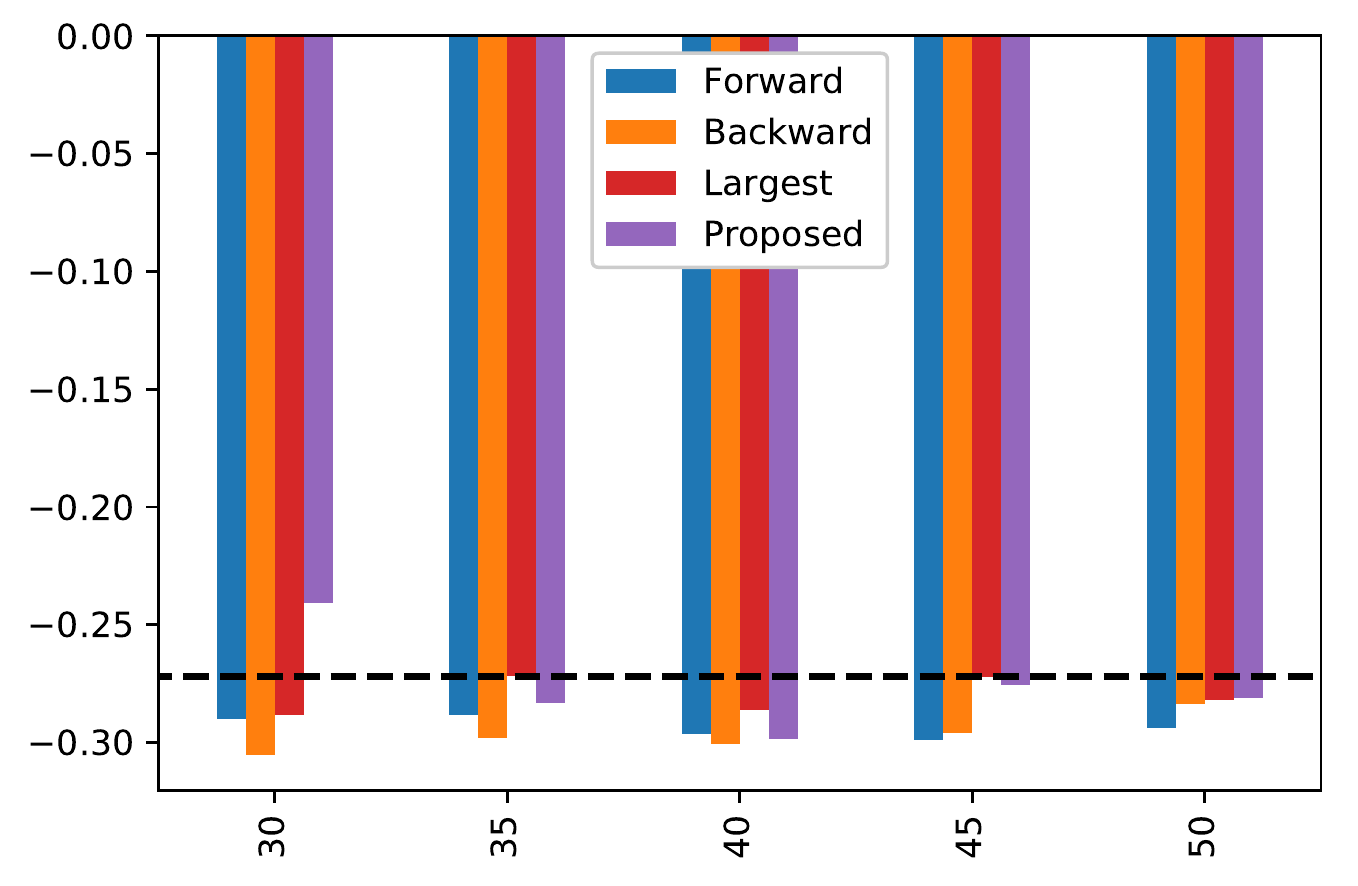}\\
\includegraphics[width=0.32\textwidth]{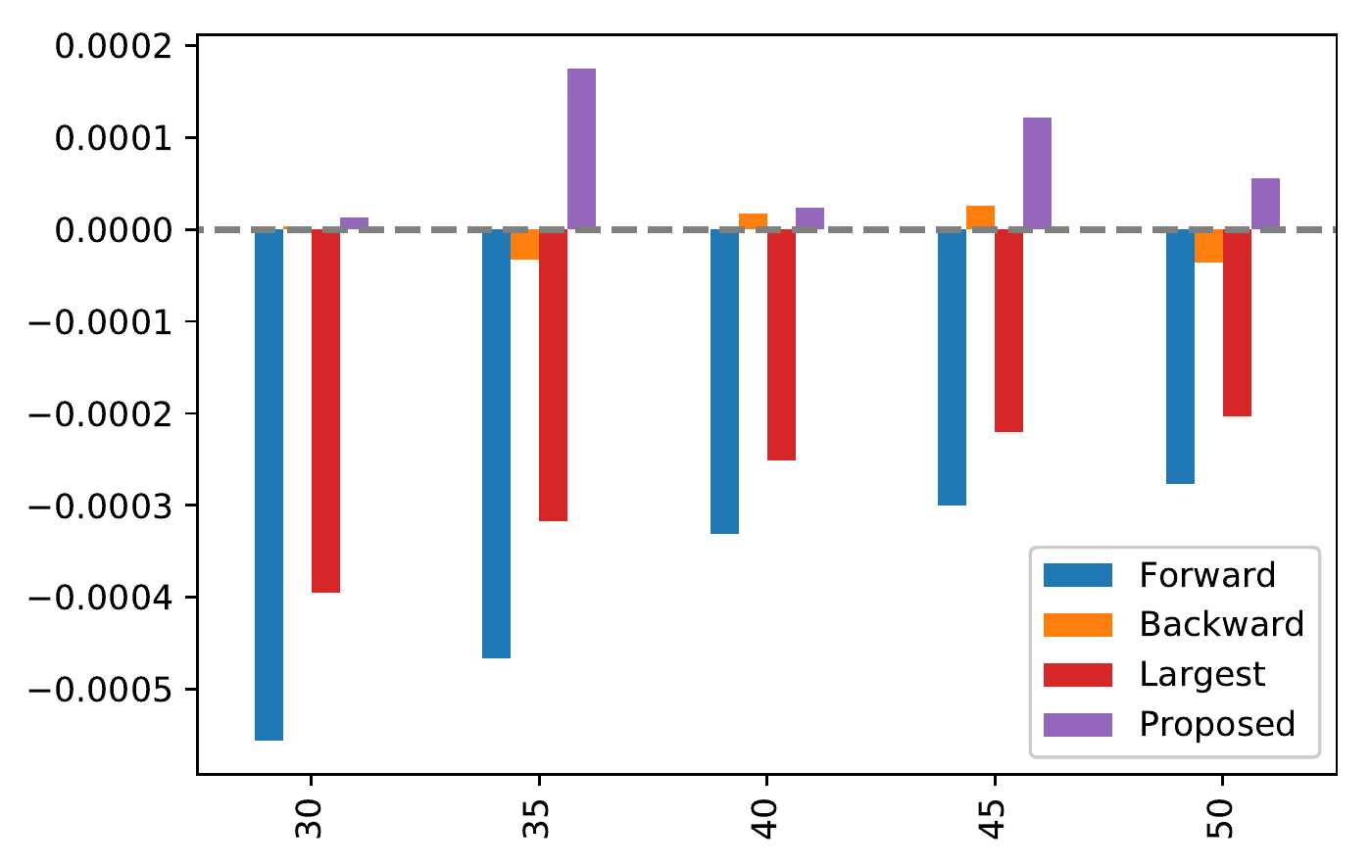}~
\includegraphics[width=0.32\textwidth]{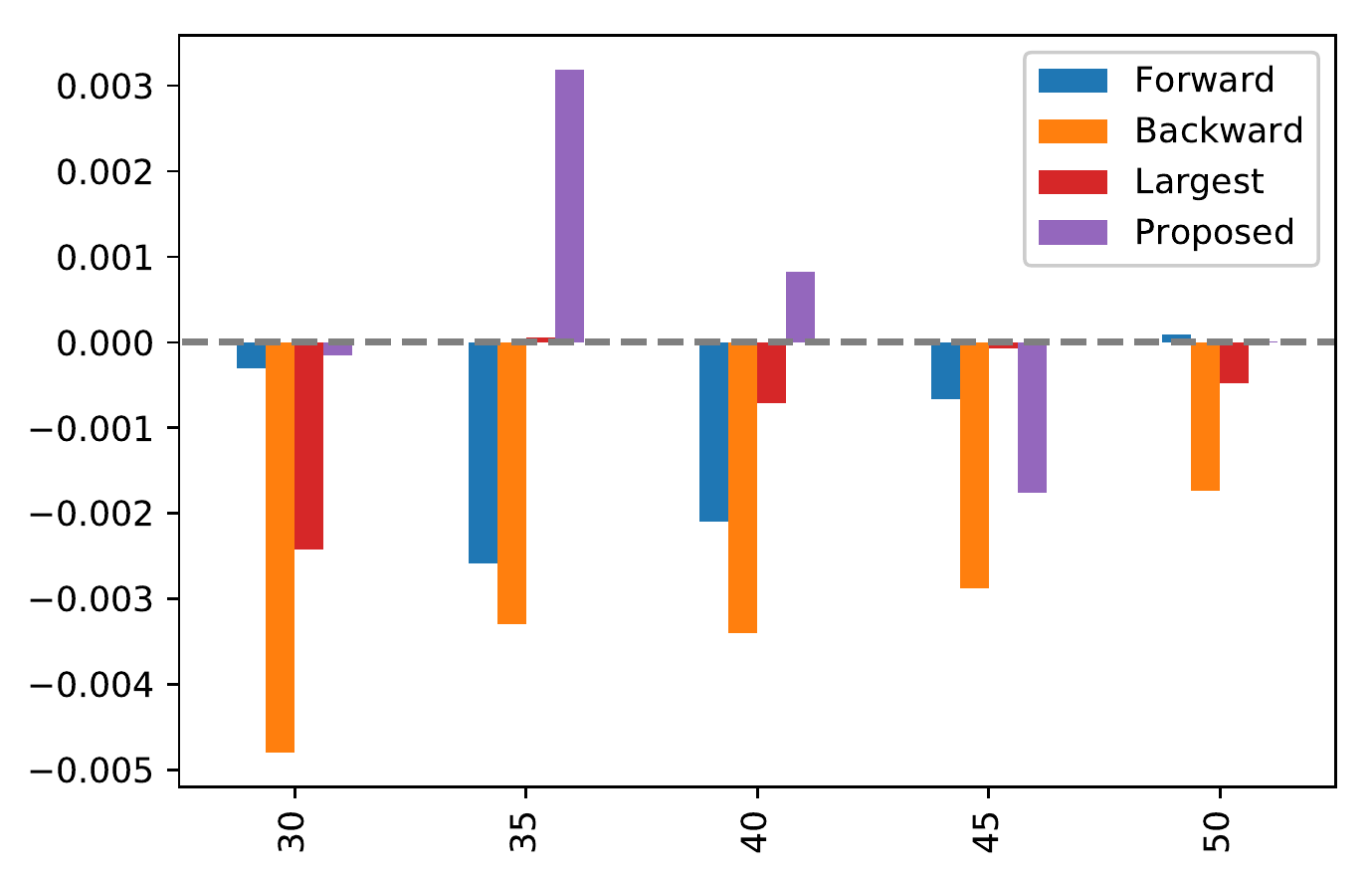}~
\includegraphics[width=0.32\textwidth]{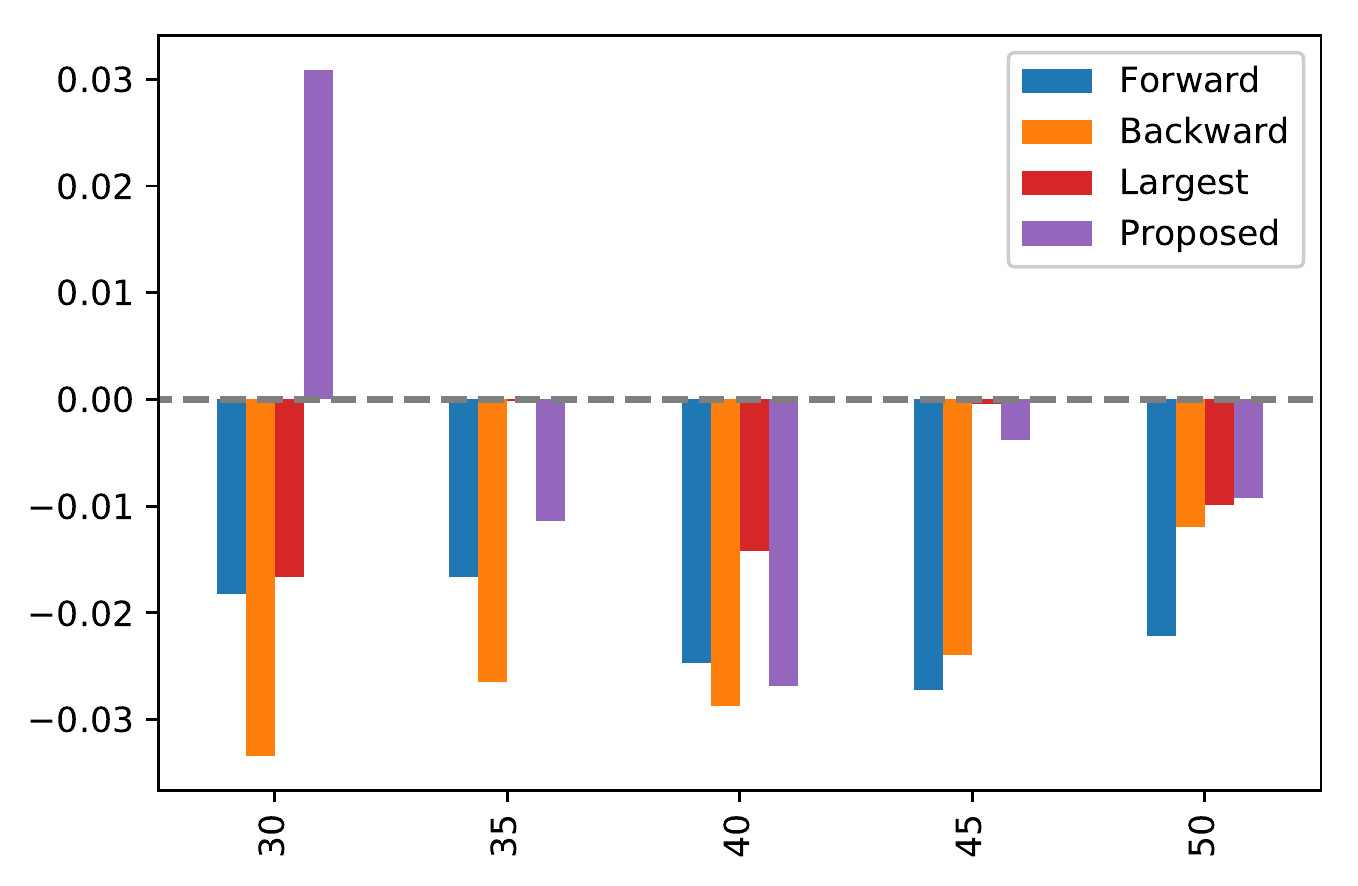}
\caption{Portfolio performance measures with various methods (colour indicators match Fig.~\ref{fig:fitting}). Left: Volatility. Middle: Sharpe Ratio. Right: Maximum Drawdown. Above: Absolute performance, with index indicated by dashed line. Below: Deviation from the index. In each plot x-axis spans various portfolio size $K$ values.}
\label{fig:stats}
\end{figure*}

Our goal is to track the index as accurately as possible. Therefore the most direct performance measure is (testing time) mean squared error (MSE) between index and partial replication portfolio performance. Considering that we use daily log-return data, the MSE is usually at the scale of $10^{-6}$, thus we report percentage error (PE) instead.

Apart from PE, we also report compute three other measures of interest: (i) Volatility of portfolio return, (ii) Sharpe ratio \cite{Sharpe94sharpe}, and (iii) Maximum Drawdown. These measures are defined as follows:

\begin{description}
\item[Volatility of portfolio return] is defined as the standard deviation of returns computed on every rebalance date. For volatility, we have run backtesting for $10$ years with quarterly rebalance, thus we have $40$ return values, and we calculate the standard deviation of those $40$ values. 
\item[Sharpe ratio] is defined as $s=\overline{r}/\sigma_r$, where $\overline{r}$ is the mean return, $\sigma_r$ is the standard deviation over that period (volatility). Thus the Sharpe ratio $s$ encodes a tradeoff of return and stability. 
\item[Maximum Drawdown] (MDD) is the measure of decline from peak during a specific period of investment: $\operatorname{MDD}=(V_{t}-V_{p})/V_{p}$, where $V_{t}$ and $V_{p}$ stand for trough and peak values, respectively. 
\end{description}

From the point of view of measuring a portfolio's performance we prefer lower, higher, and lower numbers respectively for these measures. However, our primary goal is to track the benchmark index, therefore they are of secondary importance to index tracking accuracy. For example, in cases where the index is volatile, we prefer to predict a high-volatility portfolio that accurate tracks the volatile index, rather than have a low-volatility portfolio that fails to track the index. We expect a good index tracker to match the volatility and Sharpe ratio of the index, rather than producing a low-volatility portfolio as maybe desired in some other applications.  However, for a given index-tracking accuracy we may prefer a low-volatility approximation. 

\subsubsection{Baselines}

We compare the proposed stochastic neural network method with three baselines that are widely used in commercial practice,

\begin{enumerate}
\item \textbf{Forward} Selection: We fit the constrained regression problem using QP,  select one stock with the largest weight value  $w_i$, record it in a selected stock list, and re-fit the model without the selected stock. We do this repeatedly until the size of the selected stock list reaches $K$. Finally, we fit the constrained regression problem using QP with these $K$ stocks only. 
\item \textbf{Backward} Selection: We fit the constrained regression problem using QP, and get rid of one stock with the smallest $w_i$ value, and re-fit the model. We do this repeatedly until the number of the remaining stocks is $K$.
\item \textbf{Largest} Market Capitalisation: On every rebalance day, we sort all stocks according to their market capitalisations, and pick the top $K$ stocks. Then we fit the constrained regression problem using QP with these $K$ stocks only.
\end{enumerate}

Note that, we can not evaluate those methods for which computational cost is too high, such as Evolutionary Algorithms, because we run backtesting using sliding windows over ten years for different choices on $K$, those methods are prohibitively expensive. We evaluate three choices of $K \in \{30,40,50\}$, corresponding to different portfolio cardinalities that may be desired in different real-world trading situations.

\subsubsection{Result Analysis}

Fig.~\ref{fig:fitting} compares all methods' index tracking performance at several partial replication portfolio sizes. From the graphs we can see that: (i) Our stochastic neural network approach has the highest tracking accuracy compared to the competitors (bottom row vs others). Note that, all methods have almost the same transaction costs, as the number of selected stocks is capped by $K$. Thus, the performance difference can be only explained by the merit. (iii) As expected, all methods improve tracking accuracy at higher portfolio size (columns), but ours outperforms competitors at each operating point. (iii) In terms of index tracking errors, portfolios can out-perform or under-perform the index (green plot above vs below zero in the lower percentage error plots). While the competing methods' tracking errors are mostly under-performance errors, our method's tracking errors mostly correspond to our portfolio out-performing the index (green PE plot above zero).

In terms of other measures of portfolio performance, i.e. volatility, Sharpe ratio, and maximum drawdown, our method provides comparable performance to the alternatives (Fig.~\ref{fig:stats}). Recall that the goal is to match the performance of the index in terms of these metrics, rather than optimise these metrics per-se. The deviation from the index is shown in Fig.~\ref{fig:stats}(bottom). We can see that our method achieves slightly better performance on Sharpe ratio (middle) and maximum drawdown (right) and slightly worse performance on volatility (left). As before, all methods tend to better approximate the index with larger portfolio size $K$.

\subsubsection{Stochasticity Analysis}

\begin{figure}[t!]
\centering
\includegraphics[width=0.49\textwidth]{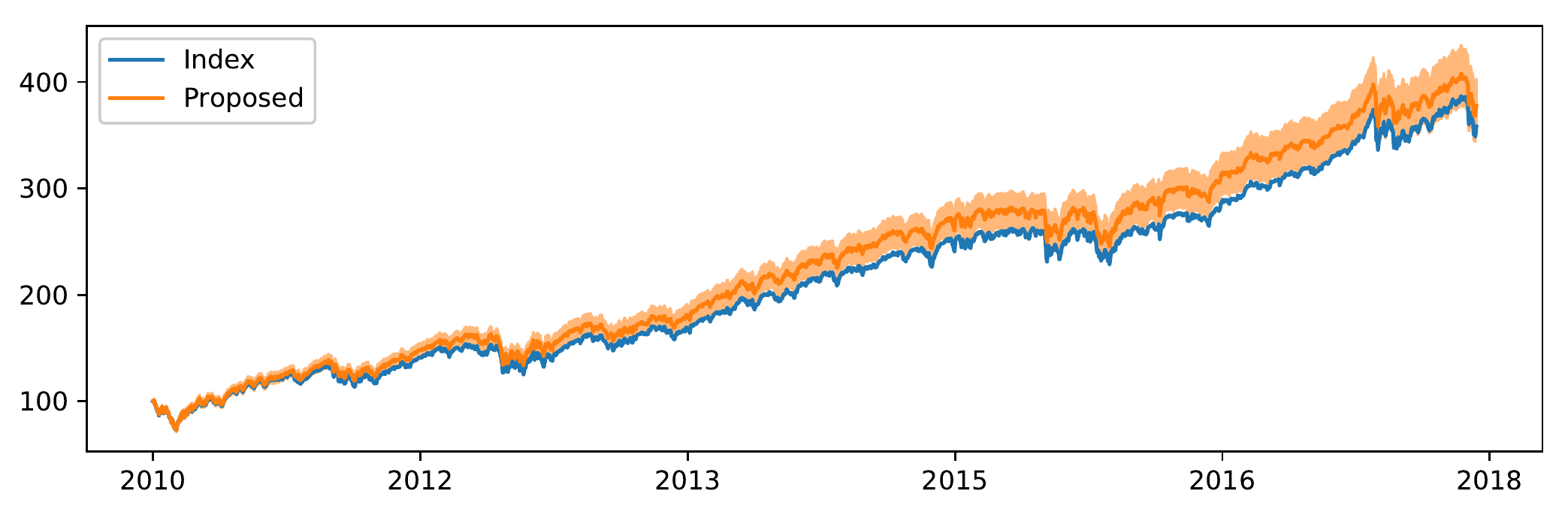}
\caption{Stochasticity Analysis. Backtesting performance of $100$ runs of our method show that it is consistently effective. Orange line is the mean partial replication index tracking performance and the shadow area covers one standard deviation. Blue line is the index.}
\label{fig:errorplot}
\end{figure}

The proposed method is non-deterministic due to the nature of the sampling process. Therefore we investigate the consistency of its performance by re-running our method $100$ times with different random seeds. The testing-stage performance in Fig.~\ref{fig:errorplot} shows that we achieve consistently good tracking performance, with the small errors being ones of outperforming the index itself.

\section{Conclusion}
\label{sec:conclusion}

We present a solution for index tracking with cardinality constraints. A novel reparametrisation method is proposed to revisit the tracking error minimisation problem and we then solve it with stochastic neural networks. Our model is simple, efficient and scalable. Detailed backtesting shows it outperforms widely used financial methods for index tracking on more than 10 years of S\&P 500 index data.\\

\noindent\textbf{Disclaimer:} All authors are faculty. Neither graduate students nor small animals were hurt while producing this paper.

\end{document}